\documentclass[12pt,epsf,amssymb]{article}
\usepackage{tabularx}
\usepackage{array}
\usepackage{graphics}
\usepackage{epsfig}
\usepackage{amsmath}
\usepackage{amssymb}
\makeatletter

\usepackage{verbatim}

\newcommand{\bea}{\begin{eqnarray}}
\newcommand{\eea}{\end{eqnarray}}
\newcommand{\beq}{\begin{equation}}
\newcommand{\eeq}{\end{equation}}
\newcommand{\gev}{{\rm GeV}}
\newcommand{\mev}{{\rm MeV}}

\newcommand{\pdir}{p\kern -5.2pt\raise 0.2ex\hbox {/}}
\newcommand{\vdir}{v\kern -5.75pt\raise 0.15ex\hbox {/}}
\newcommand{\kdir}{k\kern -5.75pt\raise 0.15ex\hbox {/}}
\newcommand{\epsdir}{\epsilon\kern -5.0pt\raise 0.15ex\hbox {/}}
\newcommand{\bvdir}{\bar{v}\kern -5.75pt\raise 0.15ex\hbox {/}}
\newcommand{\Ddir}{D\kern -7.75pt\raise 0.20ex\hbox {/}}
\newcommand{\ldir}{l\kern -5.0pt\raise 0.2ex\hbox{/}}
\newcommand{\varepsdir}{\varepsilon\kern -5.5pt\raise 0.15ex\hbox{/}}

\makeatother

\begin{document}
\thispagestyle{empty} 
\begin{flushright}
\begin{tabular}{l}
CERN-TH 99-186\\
LPT - Orsay 00/86\\
RM3-TH/00-18\\
ROMA 1304/00
\end{tabular}
\end{flushright}
\begin{center}
\vskip 3.0cm\par
{\par\centering \textbf{\LARGE   
${\mathbf{Heavy \to Light}}$  semileptonic decays of pseudoscalar mesons from lattice QCD}}\\
\vskip 0.75cm\par
{\par\centering \large  
\sc A.~Abada$^a$, D.~Becirevic$^b$, Ph.~Boucaud$^a$,\\ J.P.~Leroy$^a$, 
V.~Lubicz$^c$, F.~Mescia$^b$}
{\par\centering \vskip 0.5 cm\par}
{\sl 
$^a$ Universit\'e de Paris Sud, L.P.T. (B\^at.~210),  \\
Centre d'Orsay, 91405 Orsay-Cedex, France. \\
\vspace{.25cm}
$^b$ Dip. di Fisica, Univ. di Roma ``La Sapienza" and INFN,
Sezione di Roma,\\ 
Piazzale Aldo Moro 2, I-00185 Rome, Italy. \\
\vspace{.25cm}
$^c$ Dip. di Fisica, Univ. di Roma Tre and INFN,
Sezione di Roma III, \\
Via della Vasca Navale 84, I-00146 Rome, Italy.}\\
\vskip1.cm
 
{\vskip 0.25cm\par}
\end{center}

\vskip 0.55cm
\begin{abstract}
We have computed the form factors for $B\to \pi$ and $D\to K(\pi)$ 
semileptonic decays
on the lattice by using full non-perturbative ${\cal O}(a)$ improvement,
in the quenched approximation. 
Our results are expressed in terms of few parameters which describe the
$q^2$-dependence and normalization
of the form factors.  
\end{abstract}
\vskip 2.2 cm 
\setcounter{page}{1}
\setcounter{footnote}{0}
\setcounter{equation}{0}
%%%%%%%%%%%%%%%%%%%%%%%%%%%%%%%%%%%%%%%%
%%%%%%%%%%%%%%%%%%%%%%%%%%%%%%%%%%%%%%%%
%%%%%%%%%%%%%%%%%%%%%%%%%%%%%%%%%%%%%%%%
\noindent

\renewcommand{\thefootnote}{\arabic{footnote}}

\newpage
\setcounter{footnote}{0}
%%%%%%%%%%%  Section 1

\section{Introduction}
\setcounter{equation}{0}

In order to extract the Ca\-bibbo-Ko\-baya\-shi-Mas\-ka\-wa (CKM) matrix 
element $\vert V_{ub}\vert$ from the experimentally measured semileptonic 
branching ratio, 
$B(B^0 \to \pi^- \ell^+ \nu)$~\cite{CLEO}, 
it is essential to have a good theoretical control over the corresponding decay amplitude.
This is difficult to solve from first principles because of 
non-perturbative hadronic 
effects.

The relevant
matrix element for a generic heavy(H)$\to$light(P) decay of pseudoscalar mesons 
is parameterized as
\bea
\label{def1}
\langle P (p) \vert V_\mu \vert H(p_H)\rangle = \left( p_H + p - q {m_H^2 -
m_P^2\over q^2} \right)_\mu F_+(q^2)+ {m_H^2 - m_P^2 \over q^2} q_\mu
F_0(q^2)\;,
\eea
where $V_\mu = \bar Q\gamma_\mu q$ is the vector current involving the heavy ($Q$) and light ($q$) quark
fields, $F_{0/+}(q^2)$ are the form factors which carry the dynamical 
information on  
non-perturbative QCD, and $q=(p_H-p)$ is the momentum transferred from 
the parent to the daughter meson.

In the case of massless leptons (which is an excellent approximation for $\ell = e,\ \mu$), the 
differential decay rate reads
\bea
\label{rate}
\frac{d\Gamma}{d q^2}(H^{0} \rightarrow P^- \ell^{+} \nu_{
\ell}) \ =\ \frac{G_F^2 \vert V_{qQ}\vert ^{2}}{\; 192 \pi^3 m_H^3\; }\
\lambda^{3/2}(q^2)\ \vert F_+ (q^2) \vert ^2 \ ,
\eea
with $\lambda (q^2) = (q^2 + m_{H}^{2} - m_{P}^{2})^{2} - 
4 m_H^{2} m_{P}^{2}$, being the usual triangle function. 
To obtain the total width, one should integrate~(\ref{rate}) over the entire (large) physical
region, $0\leq q^2\leq (m_H-m_P)^2$, which requires the precise knowledge of the
normalization (say $F_{+}(0)$) and the $q^2$-dependence of the form factor.

There have been many studies devoted to the computation of these form factors. 
Various quark models have been  employed~(see {\it e.g.} 
ref.~\cite{Alain}), which in many aspects helped phenomenological 
understanding of the heavy$\to$light transitions. 
More quantitative predictions, however, can be only obtained with
approaches which are based on first QCD principles: QCD sum rules~\cite{Ruckl}~\footnote{
The most recent results for $B\to \pi$ transition, obtained by using the light cone 
QCD sum rules (LCSR), can be found in ref.~\cite{Bagan}.} and lattice 
QCD simulations~\cite{Sachrajda}.

In this paper we present new results on the form factors $F_{+/0}(q^2)$, as
obtained from our computation of the matrix element~(\ref{def1}) on the lattice. 
In order to reduce lattice artifacts,
we work on a large lattice ($24^3\times 64$) with a small spacing ($a\simeq 0.07$~fm, or
$a^{-1}=2.7(1)$~GeV),  
using the non-perturbatively ${\cal O}(a)$ improved Wilson action 
and operators~\cite{Luescher},
{\it i.e.} all leading lattice discretization effects proportional 
to the lattice spacing are absent.
Due to the limitations 
on computational resources, however, one is obliged to
make several approximations, of which the most important ones are
the use of the quenched approximation and the inability to resolve hadrons heavier than $3$~GeV.
The latter implies that the final answer for 
$B\to \pi$ transition has to be reached through an extrapolation to $m_B$, which introduces
large systematic uncertainties (which are of order $15\%$). An alternative is to discretize some effective theory,
such as NRQCD, and to compute~(\ref{def1}) from the opposite side, that is by extrapolating
from infinitely heavy
decaying mesons and eventually including higher order $1/m_H$ corrections~\cite{Hashimoto}. 
We also mention
that alternative strategies to treat $B\to \pi \ell \nu$ decay 
on the lattice have been developed in refs.~\cite{Ryan,Shigemitsu}.

The form factors are conveniently expressed in terms of three parameters 
$(c_H, \alpha_H, \beta_H)$, which define the normalization and the $q^2$ 
dependence and which follow simple scaling laws as functions of the heavy quark mass. In terms
of these parameters we have:
\bea
\label{BK}
&& \cr
&& F_+(q^2)\ =\ {c_H\ (1 - \alpha_H) \over (1 - \tilde q^2) \ (1 - \alpha_H \tilde q^2) } \;,
\cr
&& \cr
&& \cr
&& F_0(q^2)\ =\ {c_H\ (1 - \alpha_H) \over 1 - {\textstyle{\tilde q^2}\over \textstyle{\beta_H} } } \;,
\eea 
where $\tilde q^2=q^2/m_{H^\ast}^2$, and $m_{H^\ast}$ is the lightest meson that couples to
the vector current in eq.~(\ref{def1}). This form is convenient to be used in Monte Carlo
simulations for the experimental analyses of semileptonic decays. The values of the parameters
are given below, in eq.~(\ref{RES_F}).
The parameterization~(\ref{BK}), which we refer to as BK, has been introduced in ref.~\cite{Kaidalov}, 
and it encodes most of the known constraints on the form factors. This includes:
\begin{itemize}
\item[i)] the kinematical constraint, $F_+(0)=F_0(0)$;
\item[ii)] the heavy quark (meson) scaling laws predicted by the heavy 
quark effective theory (HQET), $F_+\sim \sqrt{ m_H}$ and $F_0\sim 1/\sqrt{ m_H}$,
which are applicable in the zero recoil region ($\vec q \to 0$, 
{\it i.e.} $q^2 \to q^2_{\rm max}$)~\cite{Isgur};
\item[iii)] the heavy quark scaling law predicted by the large energy effective theory (LEET) and 
explicitly realized in the LCSR framework~\footnote{This scaling law 
has been pointed out long ago by Chernyak and Zhitnitsky in ref.~\cite{Chernyak}.}, 
$F_{+/0}\sim 1/m_H^{3/2}$, applicable in the large recoil region, $q^2\to 0$~\cite{Charles,Dugan};
\item[iv)] the knowledge of the position of the first pole in the crossed channel ($q^2=m_{H^\ast}^2$), 
determining the $q^2$-dependence of $F_+(q^2)$ in the small recoil region;
\item[v)] the symmetry relation between the two form factors $F_0 = (2 E_P/m_H) F_+$, 
which is valid when the energy released to the light meson is large~\cite{Charles}~\footnote{Very
recently, it has been shown in ref.~\cite{Beneke} that this relation also holds to a high accuracy 
when radiative corrections are included.}.
\end{itemize}
In the absence of the last constraint ${\rm v})$, one additional parameter would appear in
eq.~(\ref{BK})~\cite{Kaidalov}. However, we verified that the 
relation, $F_0 = (2 E_P/m_H) F_+$, holds when applied to our data (at large recoils), 
within the statistical accuracy. Its validity has also been verified in the LCSR
approach~\cite{Khodjamirian}.

In the infinite quark mass limit, the quantities $(c_H \sqrt{m_H}, 
(1-\alpha_H) m_H, (\beta_H-1) m_H)$ 
should scale as a constant. As usual, the $1/m_H$ and $1/m_H^2$ corrections can be estimated 
from the fit with our
lattice data. This allows us to extrapolate to the $B$ meson mass. This strategy is 
employed in this paper and we call it {\sl Method I}. Another way to handle the 
extrapolations has been recently proposed by the UKQCD collaboration~\cite{UKQCD}. 
In this case the assumptions are different, namely the extrapolations  are performed 
first in the light meson  at fixed $q^2$, and then in the heavy one at a fixed value of
$v\cdot p$ where $v$ is the four-velocity of the heavy meson.
The $q^2$ dependence of the extrapolated form factors is described 
by the parameterization~(\ref{BK}). We also employed this strategy, which we call {\sl Method II}.
Both methods lead to fully consistent results. They are also in a good agreement with those of 
UKQCD~\cite{UKQCD,Maynard}.

In the form~(\ref{BK}), and by
using the two different methods, we obtain:
  
\bea 
 { \underline{\sl Method~I\ :}}  &\qquad \underline{\sl Method~II\ :}\nonumber \\ \nonumber
& \\ \nonumber
 F(0) = 0.26(5)(4)  &\qquad F(0) =
0.28(6)(5) \\ \nonumber
& \\ \nonumber
  c_B = 0.42(13)(4) &\qquad  c_B = 0.51(8)(1) \\ \label{RES_F}
& \\ \nonumber
  \alpha_B = 0.40(15)(9) &\qquad  \alpha_B = 0.45(17)^{+.06}_{-.13} \\ \nonumber
& \\ \nonumber
  \beta_B = 1.22(14)^{+.12}_{-.03} &\qquad  \beta_B = 1.20(13)^{+.15}_{-.00} \\ \nonumber
\eea
where the first errors are statistical and the second are systematic. The result for $F(0)$ 
is displayed for the reader's convenience, although it is clear that $F(0)=c_B (1-\alpha_B)$. 
We note that the results of the two methods are also in good agreement with the LCSR results 
of ref.~\cite{Khodjamirian}.

Knowing the normalization and 
the $q^2$-dependence of $F_+(q^2)$, we can integrate eq.~(\ref{rate}) to get the decay width. 
We quote:
\bea
{1\over\vert V_{ub}\vert^2}\ \Gamma (\bar B^0\to \pi^+ \ell \bar \nu) = \left(7.0\pm
2.9 \right)\ {\rm ps}^{-1}
\eea
obtained by combining the results of the two methods and adding the statistical and 
systematic errors in the quadrature. From this result, by using the measured branching ratio
$B(\bar B^0\to \pi^+ \ell \bar \nu)=(1.8\pm 0.6)\cdot 10^{-4}$~\cite{CLEO}  
and the average $B^0$ meson lifetime  $\tau_{B^0} = 1.548(32)\ {\rm ps}$~\cite{PDG}, we get
\bea
\vert V_{ub}\vert\ =\ (4.1 \pm 1.1)\cdot 10^{-3}\;.
\eea
In addition, we also study $\bar D^0\to \pi^+ \ell \bar \nu$ and 
$\bar D^0\to K^+ \ell \bar \nu$ decays and compare the results 
with the experimental data.

The remainder of this paper is organized as follows: 
in sec.~\ref{Lattice} we give the information on the lattice setup
and describe the extraction of the raw form factor 
data; in sec.~\ref{Methods} we explain the details used in
extrapolation procedures to get the physical results which are presented 
in sec.~\ref{Results} for $B\to \pi \ell \nu$ and  in sec.~\ref{ResultsD} for 
$D\to K(\pi ) \ell \nu$;
we briefly conclude in sec.~\ref{Conclusion}.

\section{Lattice Setup and Computation of the Matrix Element\label{Lattice}}

In this section we give the essential information about the lattice setup used 
in our simulation. We also sketch the strategy to compute the matrix element~(\ref{def1})
improved at ${\cal O}(a)$.

\begin{table}[h]
%\vspace*{1.cm}
\centering 
%{\underline{\sf Statistics: 200 configurations}}\\
%{\phantom{\large{l}}}\raisebox{-.1cm}{\phantom{\large{j}}}
\begin{tabular}{ccc}  \hline \hline
{\phantom{\huge{l}}}\raisebox{-.2cm}{\phantom{\Huge{j}}}
$c_{SW}$& \cite{cSW} &1.614 \\ \hline 
{\phantom{\huge{l}}}\raisebox{-.2cm}{\phantom{\Huge{j}}}
$\kappa_{heavy}\equiv \kappa_Q$& &0.1250;\ 0.1220;\ 0.1190;\ 0.1150   \\
{\phantom{\huge{l}}}\raisebox{-.2cm}{\phantom{\Huge{j}}}
$\kappa_{light}\equiv \kappa_q$& & 0.1344;\ 0.1349;\ 0.1352   \\ 
{\phantom{\huge{l}}}\raisebox{-.2cm}{\phantom{\Huge{j}}}
$a^{-1}(m_{K^*})$& & $2.7(1)$~GeV   \\
{\phantom{\huge{l}}}\raisebox{-.2cm}{\phantom{\Huge{j}}}
$\kappa_{cr}$& & $0.13585(2)$   \\
\hline
{\phantom{\huge{l}}}\raisebox{-.2cm}{\phantom{\Huge{j}}}
$Z_{V}^{(0)}$& \cite{ZVbV,cV} &0.79 \\
{\phantom{\huge{l}}}\raisebox{-.2cm}{\phantom{\Huge{j}}}
$c_{V}$& \cite{cV} &--0.09 \\
{\phantom{\huge{l}}}\raisebox{-.2cm}{\phantom{\Huge{j}}}
$b_{V}$& \cite{ZVbV,cV} &1.40 \\
 \hline \hline
\end{tabular}
%%%%%%%%%%%%%%
{\caption{\small \label{tab:1} Parameters used in this work: the Wilson hopping parameters
corresponding to the heavy ($\kappa_Q$) and light ($\kappa_q$) quark masses; $c_{SW}$ 
provides the non-perturbative ${\cal O}(a)$ improvement of the Wilson action; 
$c_V$, $b_V$, $Z_{V}^{(0)}$
ensure that the vector current is free of ${\cal O}(a)$ effects (see eq.~{\rm (\ref{def:2})}).
For each parameter we quote the reference where the quantity has been computed
non-perturbatively. The critical hopping parameter ($\kappa_{cr}$) and 
the inverse lattice spacing ($a^{-1}$) are
fixed as explained in ref.~{\rm \cite{light}}. }}
\end{table}
The results presented in this paper are obtained from a simulation on 
a lattice of  size $24^3 \times 64$, using a sample of $200$ independent gauge field 
configurations  generated, at 
$\beta = 6.2$, in the quenched approximation. 
The values of the Wilson hopping parameters corresponding to the heavy 
and light quarks are listed in tab.~\ref{tab:1}. In the same table, we 
 give the values of the other parameters necessary for the ${\cal O}(a)$ improvement 
 which we implement in this work (note that all the parameters are determined 
non-perturbatively). The statistical errors are estimated 
by using the standard jackknife procedure, by decimating each time 5 configurations from the 
whole sample.
In our previous papers~\cite{light,heavy}, we discussed the mass spectrum and decay constants 
using the ${\cal O}(a)$ improvement as obtained on a subset of 100 configurations~\footnote{
The values of the heavy-light decay constants 
were updated in refs.~\cite{Pisa,Bangalore}.}.
The updated values of
our heavy-light pseudoscalar ($H_d$) and vector  ($H^\ast_d$) meson masses in physical units, 
obtained after extrapolating to the average $up-down$ quark mass (see ref.~\cite{heavy} for
details) are:
\bea
\label{masses}
M_{H_d} &=& \big\{\;  0.642(2),\; 0.739(2),\;  0.829(2),\;  0.942(3) \;  \big\} \cr
m_{H_d} [\gev ]&=& \big\{\;  1.74(7),\; 2.01(8),\;  2.25(9),\;  2.56(10) \;  \big\} \cr
&&\cr
M_{H^\ast_d} &=& \big\{\;  0.684(3),\; 0.775(3),\;  0.860(3),\;  0.968(4) \;  \big\} \cr
m_{H^\ast_d} [\gev ]&=& \big\{\;  1.86(7),\;  2.10(8),\;  2.33(9),\;  2.63(11) \;   \big\} \;.
\eea
As in our previous publications, we denote the meson masses in lattice units by
capital letters, whereas the small case letters are used for the same masses in physical units, {\it e.g.}
$M_H = m_H a$.

To compute the matrix element~(\ref{def1}) on the lattice with 
improved Wilson fermions, the appropriate definition of the vector current reads~\cite{ZVbV}
\bea
\label{def:2}
\hat V_\mu = Z_V^{(0)}  \, \Bigl( 1+ \frac{b_V}{2} (a m_Q+ a m_q) \Bigr) \, 
\Bigl[ \bar Q \gamma_\mu q\ +\ a c_V\cdot i\partial_\nu \bar Q \sigma_{\mu\nu} q 
\Bigr] 
\eea 
with $\sigma_{\mu \nu}=(i/2)[\gamma_\mu,\gamma_\nu ]$, and where the renormalization constants $Z_V^{(0)}$ and $b_V$, as well as 
the subtraction constant $c_V$, are known quantities (in tab.~\ref{tab:1} we also 
quote the reference in which a given quantity 
has been computed non-perturbatively). However, since we deal with the heavy-light 
operator, one can suspect that higher order terms, $\propto (am)^n$ ($n\geq 2$), 
may be important. Their cancellation at tree level can be achieved by multiplying $Z_V$ by 
the KLM factor~\cite{KLM} which we modify to include ${\cal O}(a)$ improvement~\cite{heavy}, {\it i.e.}
\bea \label{klm}
Z_{V}^{\rm KLM}= Z_{V}^{(0)} {\sqrt{ 1 + am_q} \sqrt{ 1 + am_Q}\over 1 + a\bar m}
\left( 1 + b_V a \bar m \right)\; ,
\eea 
where $a m_{q\{Q\}} = (1/\kappa_{q\{Q\}} - 1/\kappa_{cr})/2$ and $\bar m = (m_q + m_Q)/2$.

To access the matrix element~(\ref{def1}) on the lattice, one computes 
the following three- and two-point correlation functions
\bea
\label{3pts}
{\cal C}^{(3)}_{\mu} (t_x, t_y;\vec q,\vec p_H) &=& 
\int d \vec x d \vec y\ e^{i (\vec q\cdot \vec y - \vec p_H\cdot \vec x )} 
\langle   P(0)\hat V_\mu(y) H^\dagger(x)  \rangle\, , \cr
&&\cr
{\cal C}_{JJ}^{(2)}(t) &=& \int d \vec x \ e^{i\vec p\cdot \vec x } 
 \langle   J({\vec x}, t) 
J^{\dagger}(0) \rangle\, ,
\eea
where $H=\bar q i\gamma_5 Q$ and $P=\bar q i\gamma_5 q$ are the source operators of the 
heavy-light and light-light pseudoscalar mesons, respectively, and $J$ denotes any of the two ($H$ or
$P$). On the temporal axis in ${\cal C}^{(3)}_{\mu}$, we fixed
$H(x)$ to $t_x=27$, and the improved 
renormalized  vector current
$\hat V_\mu(y)$ defined in eq.~(\ref{def:2}) is free ($0\leq t_y\leq 63$).
$\vec p_H$ is the three-momentum given to the decaying meson, 
and $\vec q$ denotes the momentum transferred to the daughter meson 
($\vec q=\vec p_H - \vec p$). In this work we consider the following 13 kinematical configurations:
\bea
\label{momenta}
{\underline{\vec p_H =\left( 0,0,0\right)}}&&
\vec q = \big\{ \left( 0,0,0\right); \left( 1,0,0\right) ;\left( 1,1,0\right) \big\}\\ \nonumber
&&\\ \nonumber
{\underline{\vec p_H =\left( 1,0,0\right)}}&&
\vec q = \big\{ \left( 0,0,0\right); \left( 1,0,0\right); \left( 0,1,0\right);
\left( 2,0,0\right);\left( 1,1,0\right);\left( 1,1,1\right);\left(
2,1,0\right) \big\}\\ \nonumber
&&\cr
{\underline{\vec p_H =\left( 1,1,0\right)}}
&& \vec q = \big\{ \left( 1,0,0\right); \left(
1,1,0\right);\left( 2,1,0\right)  \big\}
  \eea
where each component is given in units of elementary momentum on 
the lattice $(2 \pi/La)$. 
As usual, to get a better statistical quality of the signal, we averaged over 
the equivalent configurations, {\it i.e.} the ones that can be obtained
by applying the cubic lattice symmetry, as well as the parity and 
charge conjugation transformations.

From the asymptotic behavior of the three-point correlation function
\bea \label{asym3}
{\cal C}^{(3)}_\mu (t_x,t_y;\vec q, \vec p_H) \stackrel{t_x\gg t_y \gg 0}{\longrightarrow}\, 
{\sqrt{{\cal Z}_{P}} \over 2 E_{P}} \ e^{- E_{P} t_y} \cdot \langle P (\vec p_H-\vec q)\vert \hat V_\mu(0)
\vert H(\vec p_H)\rangle \cdot 
{\sqrt{{\cal Z}_{H}} \over 2 E_{H}} \ e^{-
E_{H}(t_x - t_y)}\,,
\eea
it is obvious that the removal of the exponential factors can be achieved by considering the ratio
\bea
\label{ratio}
R_\mu(t_y) = 
{{\cal C}^{(3)}_\mu (t_x,t_y;\vec q, \vec p_H)\over \quad {\cal C}^{(2)}_{PP}(t_y,\vec p_H - \vec q) 
\, \ {\cal C}^{(2)}_{HH}(t_x - t_y,\vec p_H)  \quad } \cdot 
{ \sqrt{{\cal Z}_{H}} \sqrt{{\cal Z}_{P}} }\;,
\eea
where $\sqrt{{\cal Z}_{H}} = \vert\langle 0\vert \bar q i\gamma_5 Q\vert H \rangle\vert$ and
$\sqrt{{\cal Z}_{P}} = \vert\langle 0\vert \bar q i\gamma_5 q\vert P \rangle\vert$, are obtained from
the fit with asymptotic behavior of the two-point correlation functions.
\bea
\label{asym2}
{\cal C}_{JJ}^{(2)}(t) \, \stackrel{t\gg 0}{\longrightarrow}\, {{\cal{Z}}_J
 \over 2 E_J}e^{- E_J t} \ .
\eea
When the operators in the ratio~(\ref{ratio}) are sufficiently separated, one observes 
the stable signal (plateau), which is the desired matrix
element:
\bea
\label{matel}
R_\mu(t_y) \stackrel{t_x\gg t_y \gg 0}{\longrightarrow} 
\langle P (p) \vert \hat V_\mu \vert H(p_H)\rangle\;.
\eea
%%%%%%%%%%%%%%%%%%%%%%%%%%%%%%%%%%%%%%%%%%%%%%%%%%%%%%%%%%%
\begin{figure}
\vspace*{-5.cm}
\begin{center}
\begin{tabular}{@{\hspace{-0.7cm}}c}
\epsfxsize16.0cm\epsffile{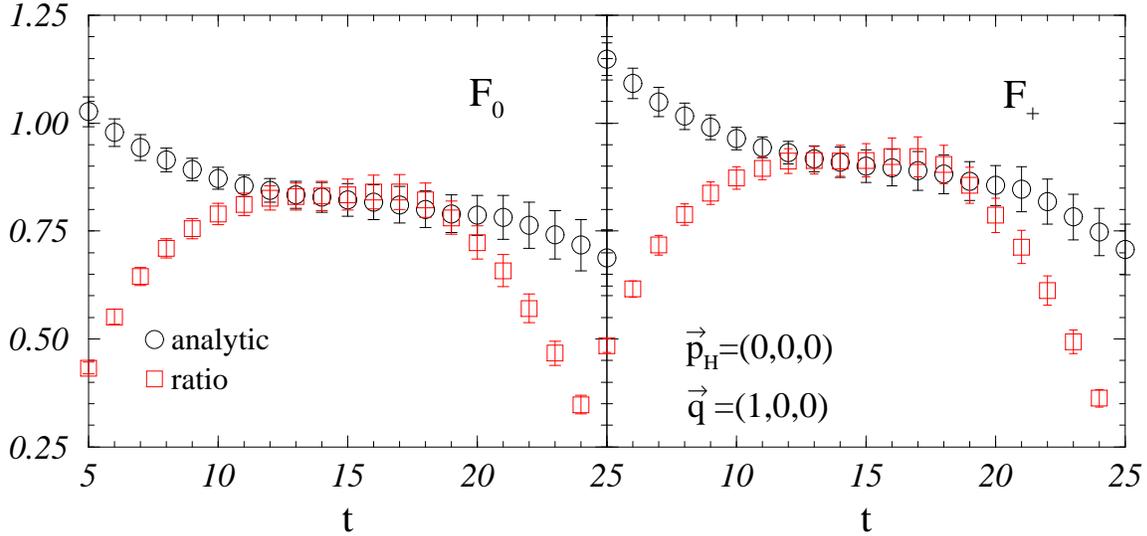}   \\
\end{tabular}
%%%%%%%%%%%%%%%%%%%%%%%%%%%%%%%%%%%%%%%%%%%%%%%%%%%%%%%%%%%%%%%%%%
\caption{\label{fig1}{\small Plateaus of the two form factors
for the specific combination of the heavy and light quark masses (corresponding
to $\kappa_Q = 0.1220$ and $\kappa_q=0.1349$). The form factors are extracted from the fit 
in $t\in [12,15]$.}}
%%%%%%%%%%%%%%%%%%%%%%%%%%%%%%%%%%%%%%%%%%%%%%%%%%%%%%%%%%%%%%%%%%
\end{center}
\end{figure}
%%%%%%%%%%%%%%%%%%%%%%%%%%%%%%%%%%%%%%%%%%%%%%%%%%%%%%%%%%%
The plateaus are typically found for $12 \leq t_y\leq 15$. Fitting $R_\mu(t_y)$ to a constant
in that interval, we extract the matrix element~(\ref{def1}). The other possibility is to 
express $R_\mu(t_y)$ in terms of form factors for each time-slice, $t\equiv t_y$, and then to 
fit each form factor to a constant in the common plateau interval.  
As an important cross check of this procedure, we employ the following two methods~\cite{ELC}:
\begin{itemize}
\item[--] {\sf Ratio method} consists in forming the ratio~(\ref{ratio}) of the 
correlation functions computed on the lattice, and using ${\cal Z}_{P}$ and 
${\cal Z}_{H}$ obtained from the fit with the corresponding two-point correlation functions
in the same time-interval as the one at which the plateau has been attained. 
\item[--] {\sf Analytic method} refers to the procedure in which we replace the two-point correlation functions
in eq.~(\ref{ratio}) by their  asymptotic expressions~(\ref{asym2}) and 
 ${\cal Z}_{P}$ and ${\cal Z}_{H}$ are extracted from 
the separate study of the two-point correlation functions. In this case we also take into
account the symmetry of ${\cal C}_{JJ}^{(2)}(t)$ with respect to the time inversion, 
$t \to (64 - t)$. 
In this work we use the latticized free boson dispersion relation:
\bea
{\rm sinh}^2\left( {E_J(\vec p)\over 2} \right) = {\rm sinh}^2\left( {M_J\over 2}
\right) + {\rm sin}^2\left( {\vec p\over 2} \right)\ ,
\eea
which appears to be appropriate in describing our data~\cite{light,heavy}. 
\end{itemize}
The differences in the results obtained by
using the ratio and analytic methods are always smaller than 
the statistical errors. In fig.~\ref{fig1}, we illustrate the form factors $F_{0/+}(t)$ as functions of
time $t\equiv t_y$,  
obtained from
the ratio~(\ref{ratio}) by using both the ratio and the analytic methods. From now on, 
our central results will always be those obtained by using the ratio method. The difference 
between  that result and the one we get by applying the analytic method, 
will be added to the systematic uncertainties. We have also checked whether the
results change after replacing $|\vec p|\to 2 {\rm sin}(|\vec p|/2)$. This difference,
which is an ${\cal O}(a^2)$ effect, is around $1\%$ and we 
neglect it. This is in agreement 
with findings of ref.~\cite{gluons} where this source of discretization errors 
was studied in great details for the gluon propagator and the 
triple gluon vertex. These errors, although important for large momentum
injections, were 
shown to be completely negligible for the small momenta (such as the ones
 considered in this paper~(\ref{momenta}).) This indication makes us confident 
 that this source of systematic errors is under control.
%%%%%%%%%%%%%%%%%%%%%%%%%%%%%%%%%%%%%%%%%%%%%%%%%%%%%%%%%%%
\begin{figure}
\vspace*{-4.5cm}
\begin{center}
\begin{tabular}{@{\hspace{.7cm}}c}
\epsfxsize17.cm\epsffile{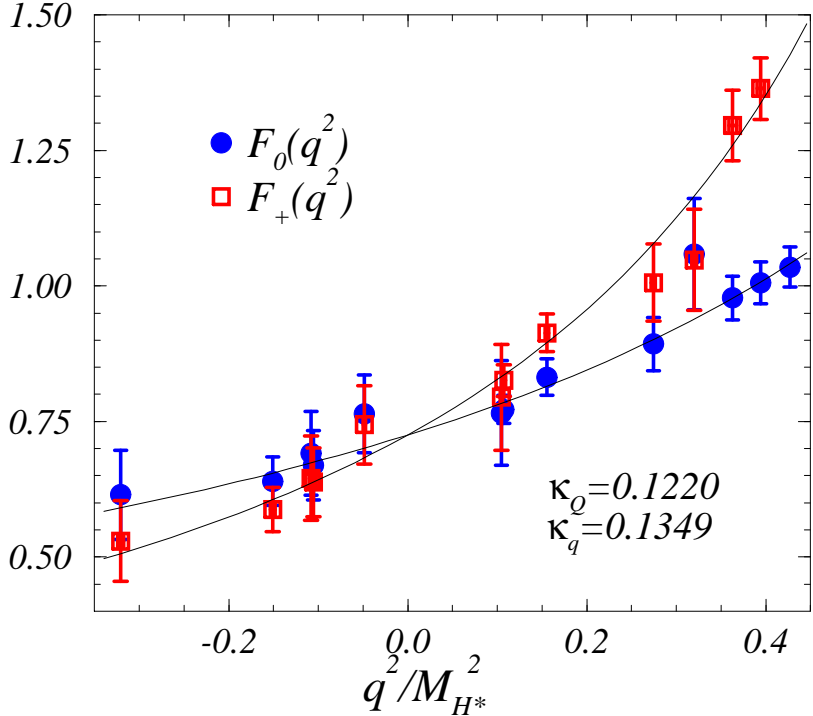}   \\
\end{tabular}
\vspace*{-1.2cm}
%%%%%%%%%%%%%%%%%%%%%%%%%%%%%%%%%%%%%%%%%%%%%%%%%%%%%%%%%%%%%%%%%%
\caption{\label{fig2}{\small The form factors $F_{0/+}(q^2)$ for 
$\kappa_q=0.1349$ and $\kappa_Q = 0.1220$, plotted as functions of $q^2/M_V^2$,
where  $M_V$ is the mass of the lightest vector meson exchanged in the $t$-channel. 
The lines show the fit to the parameterization~(\ref{BK}).}}
%%%%%%%%%%%%%%%%%%%%%%%%%%%%%%%%%%%%%%%%%%%%%%%%%%%%%%%%%%%%%%%%%%
\end{center}
\end{figure}
%%%%%%%%%%%%%%%%%%%%%%%%%%%%%%%%%%%%%%%%%%%%%%%%%%%%%%%%%%%

\section{Getting to the physical results: Extrapolations\label{Methods}}

Once we computed the form factors for all 13 momentum combinations~(\ref{momenta}), we need to 
address the question of the mass extrapolations. Firstly, since our light meson 
masses are in the range $0.54~\gev \lesssim m_P \lesssim 0.84~\gev$, a small extrapolation to the 
physical kaon and a longer one to the pion mass are needed. Secondly, the extrapolation
 in the heavy quark mass is needed, if we are to reach the physical results relevant for
 $B\to \pi$ decay. These extrapolations are closely related to the $q^2$ fit of the form factors
 because both extrapolations, in light and heavy mesons, obviously enlarge the
physical region accessible from the semileptonic decay, $q^2\in [0, (m_H-m_P)^2]$.
Before discussing the extrapolations, we 
list in tab.~\ref{tab:PAR} the BK-parameters~(\ref{BK}) 
resulting from the fit to our data for each combination of 
the heavy and light quarks. One such a fit is
shown in fig.~\ref{fig2}.
\begin{table}[h]
\centering
\begin{tabular}{|c|c|c|c||c|} \hline
{\phantom{\huge{l}}}\raisebox{-.2cm}{\phantom{\Huge{j}}}
\hspace{-1.5mm}$\kappa_Q$--$\kappa_q$ & { $c_H$} & { $\alpha_H$} & $\beta_H$ & { $F^{H\to P}(0)$} \\ \hline \hline
{\phantom{\Large{l}}}\raisebox{.2cm}{\phantom{\Large{j}}}
{\hspace{-1.5mm}$1250-1344$\hspace{1.5mm}} & 1.23(12) & 0.30(7) & 1.38(11) & 0.87(2) \\
{ $1250-1349$} & 1.09(13) & 0.29(9) & 1.37(12) & 0.77(3) \\
{\phantom{\Large{l}}}\raisebox{-.2cm}{\phantom{\Large{j}}}
{\hspace{-1.5mm}$1250-1352$\hspace{1.5mm}} & 0.99(13) & 0.29(11) & 1.33(13) & 0.70(4) \\ \hline
\hline
{\phantom{\Large{l}}}\raisebox{.2cm}{\phantom{\Large{j}}}
{\hspace{-1.5mm}$1220-1344$\hspace{1.5mm}}  & 1.13(12) & 0.28(9) & 1.42(13) & 0.82(2) \\
{ $1220-1349$} & 0.99(12) & 0.26(10) & 1.41(13) & 0.73(3) \\
{\phantom{\Large{l}}}\raisebox{-.2cm}{\phantom{\Large{j}}}
{\hspace{-1.5mm}$1220-1352$\hspace{1.5mm}}  & 0.88(11) & 0.26(12) & 1.35(14) & 0.65(4) \\ \hline
  \hline
{\phantom{\Large{l}}}\raisebox{.2cm}{\phantom{\Large{j}}}
{\hspace{-1.5mm}$1190-1344$\hspace{1.5mm}} & 1.08(12) & 0.28(9) & 1.43(13) & 0.78(3) \\
{ $1190-1349$} & 0.93(11) & 0.26(11) & 1.42(14) & 0.69(3)  \\
{\phantom{\Large{l}}}\raisebox{-.2cm}{\phantom{\Large{j}}}
{\hspace{-1.5mm}$1190-1352$\hspace{1.5mm}} & 0.83(10) & 0.27(11) & 1.33(12) & 0.61(4) \\ \hline
\hline
{\phantom{\Large{l}}}\raisebox{.2cm}{\phantom{\Large{j}}}
{\hspace{-1.5mm}$1150-1344$\hspace{1.5mm}} & 1.02(11) & 0.28(10) & 1.43(14) & 0.73(3) \\
{ $1150-1349$} & 0.88(10) & 0.26(12) & 1.42(14) & 0.65(4) \\
{\phantom{\Large{l}}}\raisebox{-.2cm}{\phantom{\Large{j}}}
{\hspace{-1.5mm}$1150-1352$\hspace{1.5mm}} & 0.77(9) & 0.25(12) & 1.35(13) & 0.58(4) \\
\hline \end{tabular}
\vspace*{0.7cm}
\caption{\small The results of the fit of our form factor data to the parameterization
given in eq.~(\ref{BK}), for all combinations of $\kappa_Q$--$\kappa_q$.}
%\end{center}
\label{tab:PAR}
\end{table}

\subsection{Method I: Extrapolation of the parameters}

%%%%%%%%%%%%%%%%%%%%%%%%%%%%%%%%%%%%%%%%%%%%%%%%%%%%%%%%%%%
%%%%%%%%%%%%%%%%%%%%%%%%%%%%%%%%%%%%%%%%%%%%%%%%%%%%%%%%%%%
\begin{figure}
%\vspace*{-3.cm}
\begin{center}
\begin{tabular}{@{\hspace{.1cm}}c}
\epsfxsize17.0cm\epsffile{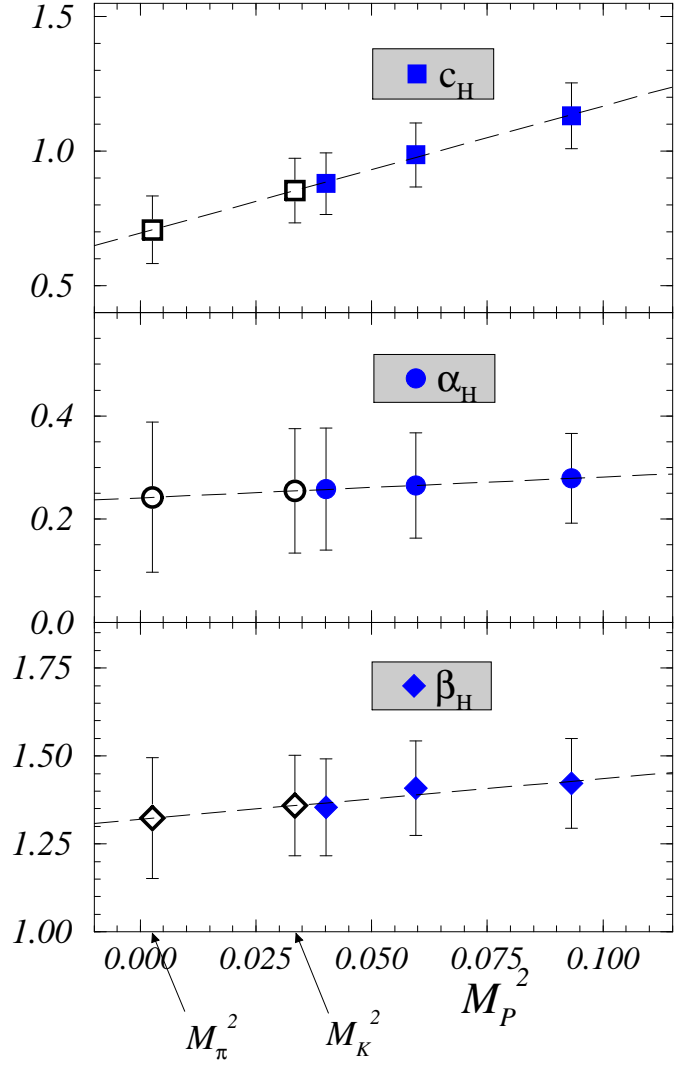}   \\
\end{tabular}
%%%%%%%%%%%%%%%%%%%%%%%%%%%%%%%%%%%%%%%%%%%%%%%%%%%%%%%%%%%%%%%%%%
%\vspace*{-1.25cm}
\caption{\label{fig6}{\small Extrapolation of each of the parameters 
$(c_H,\ \alpha_H, \ \beta_H)$ to the final kaon/pion. The plot is made for the heavy quark  
corresponding to $\kappa_Q = 0.1220$.}}
%%%%%%%%%%%%%%%%%%%%%%%%%%%%%%%%%%%%%%%%%%%%%%%%%%%%%%%%%%%%%%%%%%
\end{center}
\end{figure}
%%%%%%%%%%%%%%%%%%%%%%%%%%%%%%%%%%%%%%%%%%%%%%%%%%%%%%%%%%%
%%%%%%%%%%%%%%%%%%%%%%%%%%%%%%%%%%%%%%%%%%%%%%%%%%%%%%%%%%%
The parameters $\varphi_i = (c_H,\ \alpha_H, \ \beta_H)$, for each pair of 
the heavy and light hopping  parameters (see tab.~\ref{tab:PAR}) give us information 
on the normalization and the $q^2$-dependences of both form factors.
For fixed heavy quark, one can apply the lattice plane method~\cite{Allton} to 
extrapolate to the case when the final meson is either a kaon or a pion. This essentially 
means a linear fit in the light quark mass ($M_P^2\propto m_q$) for each parameter:
\bea \label{planes}
\varphi_i = A_i + B_i\ M_P^2\;,
\eea
and then an extrapolation to $M_\pi^2$, determined in the $(M_P^2, M_V)$ 
plane by fixing the ratio $M_V/M_P$ to the physical point $m_\rho/m_\pi$. 
Similarly, the value of $M_K^2$ is determined from the physical $m_{K^\ast}/m_K$ ratio. 
The result of that extrapolation is illustrated in
fig.~\ref{fig6}, and the extrapolated parameters are listed in tab.~\ref{tab:4}.
\begin{table}
\begin{center}
\hspace*{-3mm}\begin{tabular}{|c||ccc|c||ccc|c|} 
\multicolumn{1}{c}{}&\multicolumn{4}{c}{$\overbrace{\hspace*{7cm}}^{\textstyle{ Final\ meson\ -\ kaon}}$}
&\multicolumn{4}{c}{$\overbrace{\hspace*{7cm}}^{\textstyle{ Final\ meson\ -\ pion}}$}\\
  \hline
\hspace*{-4mm}{\phantom{\huge{l}}}\raisebox{-.2cm}{\phantom{\Huge{j}}}
 $\kappa_Q$ & $c_H$& $\alpha_H$& $\beta_H$ & $F (0)$ &
 $c_H$& $\alpha_H$& $\beta_H$ & $F (0)$ \\ 
\hline   \hline 
{\phantom{\Large{l}}}\raisebox{.2cm}{\phantom{\Large{j}}}
\hspace*{-4mm}0.1250 & 0.96(14)& 0.29(11)& 1.33(13)& 0.69(4) & 
         0.82(16)& 0.29(14)& 1.30(17)& 0.59(6)\\
{\phantom{\Large{l}}}\raisebox{.2cm}{\phantom{\Large{j}}}
\hspace*{-4mm}0.1220 & 0.85(12)& 0.25(12)& 1.36(14)& 0.64(4) & 
         0.71(12)& 0.24(15)& 1.32(17)& 0.55(6)\\
{\phantom{\Large{l}}}\raisebox{.2cm}{\phantom{\Large{j}}}
\hspace*{-4mm}0.1190 & 0.80(11)& 0.26(12)& 1.34(13)& 0.60(4) & 
         0.66(11)& 0.25(14)& 1.29(15)& 0.51(6)\\
{\phantom{\Large{l}}}\raisebox{.2cm}{\phantom{\Large{j}}}
\hspace*{-4mm}0.1150 & 0.74(10)& 0.24(13)& 1.35(13)& 0.57(4) & 
         0.60(10)& 0.23(15)& 1.31(15)& 0.48(6)\\
  \hline 
\end{tabular}
%%%%%%%%%%%%%%%%%%%%%%%%%%%%%%%%%%%%%%%%%%%%%%%%%%%%%%%%%%%%%%%%%%
\caption{\label{tab:4}{\small Values of the BK-parameters after extrapolating 
to the final $\pi$ and $K$ meson. The extrapolation of the form factor at $q^2=0$ provides a cross
check of the extrapolation.}}
\end{center}
\end{table}
%%%%%%%%%%%%%%%%%%%%%%%%%%%%%%%%%%%%%%%%%%%%%%%%%%%%%%%%%%%%%%%%%%
%%%%%%%%%%%%%%%%%%%%%%%%%%%%%%%%%%%%%%%%%%%%%%%%%%%%%%%%%%%%%%%%%%
%%%%%%%%%%%%%%%%%%%%%%%%%%%%%%%%%%%%%%%%%%%%%%%%%%%%%%%%%%%%%%%%%%
The form factor at $q^2=0$ is just the combination
$c_H(1-\alpha_H)$. We note that whether we extrapolate the parameters $c_H$ and $\alpha_H$ to the
final
pion (kaon) separately and then combine them to $F(0)$, or we combine the parameters to obtain $F(0)$ and
then extrapolate to the pion (kaon), the result remains the same. The latter number is also
given in tab.~\ref{tab:4}.

\begin{figure}
%\vspace*{-3.cm}
\begin{center}
\begin{tabular}{@{\hspace{.1cm}}c}
\epsfxsize17.0cm\epsffile{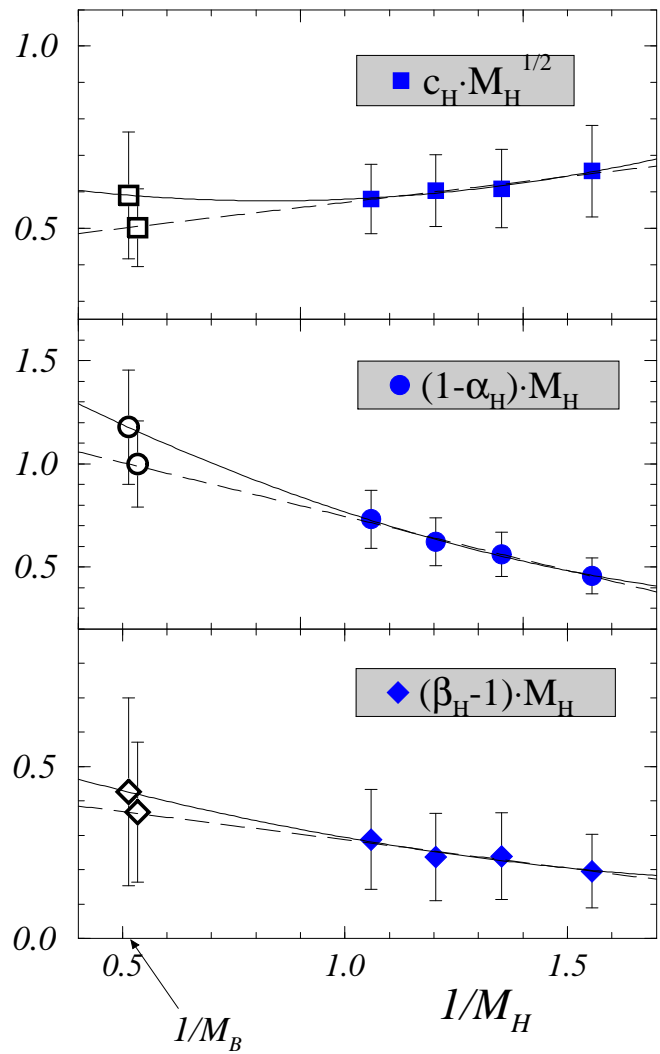}   \\
\end{tabular}
%%%%%%%%%%%%%%%%%%%%%%%%%%%%%%%%%%%%%%%%%%%%%%%%%%%%%%%%%%%%%%%%%%
%\vspace*{-1.25cm}
\caption{\label{fig7}{\small Extrapolation in the heavy meson mass. The curves represent the results 
of the quadratic (full) and linear (dashed) extrapolations.}}
%%%%%%%%%%%%%%%%%%%%%%%%%%%%%%%%%%%%%%%%%%%%%%%%%%%%%%%%%%%%%%%%%%
\end{center}
\end{figure}

After the extrapolation in the light mass, one needs to fit the parameters in the inverse 
heavy meson mass and to interpolate
to the $D$ or extrapolate to the $B$ meson mass. The quantities  which scale in the
infinitely heavy quark limit ($\phi$), are fitted as
\bea
\label{hqs}
\phi = b_0 +{ b_1\over M_H} + {b_2  \over M_H^2}\;,\quad {\rm where}\quad \phi \in 
\left\{
c_H\sqrt{M_H},\ (1-\alpha_H) M_H,\ (\beta_H -1)M_H\right\}\;,
\eea
where we neglect the logarithmic corrections.
In fig.~\ref{fig7} we show the $1/M_H$ behavior of each parameter for the case in which the final 
light meson is the pion.
%%%%%%%%%%%%%%%%%%%%%%%%%%%%%%%%%%%%%%%%%%%%%%%%%%%%%%%%%%%
%%%%%%%%%%%%%%%%%%%%%%%%%%%%%%%%%%%%%%%%%%%%%%%%%%%%%%%%%%%
%%%%%%%%%%%%%%%%%%%%%%%%%%%%%%%%%%%%%%%%%%%%%%%%%%%%%%%%%%%
Again, as a cross check, one can also fit 
$F(0) M_H^{3/2}$ to the form~(\ref{hqs}). The results for $F(0)$ extrapolated
to $M_B$ in that way is fully consistent with $F(0)$ obtained by combining 
separately the extrapolated values of $c_B$ and $\alpha_B$ to $F(0)=c_B(1-\alpha_B)$. 
As can be seen from fig.~\ref{fig7}, however, a systematic difference in the
extrapolated values of the parameters is obtained depending on whether we
perform a linear ($b_2=0$) or a quadratic ($b_2$ free) fit in the inverse
heavy meson mass. Indeed, besides the quenched approximation, this
difference represents the main systematic uncertainty present in our
calculation.
To make a 
comparison with other approaches easier, we give the values of the $1/m_H$
corrections (in physical units) that are obtained from the linear and quadratic fit with our data:
\bea
&&F(0) = {(3.1\pm 0.5)\ \gev^{3/2}\over m_H^{3/2}}  \times 
\left[ \ 1 \ - \ {(0.98\pm 0.09)\ \gev   \over m_H} \right]\,,\cr
&& \cr
&&F(0) = {(4.7\pm 1.1)\ \gev^{3/2}\over m_H^{3/2}}  \times 
\left[ \ 1 \ - \ {(2.0\pm 0.3)\ \gev   \over m_H}  \ + \ { (1.2\pm 0.1\ \gev)^2   \over m_H^2} 
\right]\,.
\eea
We observe, in passing, that the $1/m_H$ corrections are large and comparable to the ones 
which appear with the heavy-light decay
constants~\cite{heavy}.

At this point we should decide which result, from either the linear or 
quadratic fit, 
to quote as the central one.  
As a criterion for that, we test whether the result of the extrapolation
satisfies the  soft pion theorem, {\it i.e.} the Callan-Treiman 
relation applied to the case of $B\to \pi \ell \nu$~\cite{Voloshin}:~\footnote{An extensive 
high statistic study of this relation on the lattice, in the NRQCD framework, has been 
made in ref.~\cite{Hashimoto}.} 
\bea
\label{SPT}
F_0(m_B^2)\ =\ {f_B\over \ f_\pi \ }\;,
\eea
where $f_B$ and $f_\pi$ are the corresponding decay constants. The values of the ratio
$f_B/f_\pi$ will be given in the next section. 
It turns out that after the quadratic extrapolation in $1/M_H$~(\ref{hqs}), we find a better
consistency of $F_0(m_B^2)$ with $f_B/f_\pi$. 
Thus, the result of the quadratic $1/M_H$-extrapolation will be taken as our central value, 
and the difference between
the quadratically and linearly extrapolated values will be accounted for in the estimate of the 
systematic error. All the numerical results of the extrapolations will be given in next section,
where we discuss our physical results.

\subsection{Method II: UKQCD}

The second method that we use in this paper is the one proposed by the UKQCD collaboration 
in ref.~\cite{UKQCD}. As in the previous subsection, the starting point  
are the results given in tab.~\ref{tab:PAR}. For a given value of the heavy quark 
mass, one extrapolates the form factors to the final pion (kaon) by 
using~(\ref{planes}). In order to separate the intrinsic light quark mass dependence of the form factors from
the one that arises from the variation in $q^2$, the 
extrapolation in the light quark mass should be made at fixed $q^2$~\cite{UKQCD}. In other words,
 we consider the extrapolation 
to the pion for a constant $v\cdot p$, which is defined as~\footnote{Note that the light meson mass 
in eq.~(\ref{fixvp}) is the physical pion mass.}
\bea \label{fixvp}
v\cdot p = {M_{H_d}^2 + M_\pi^2 -q^2\over 2 M_{H_d}}\;,
\eea
where $v$ is the four-velocity of the heavy meson and $p$ 
is the momentum of the light meson. %%%%%%%%%%%%%%%%%%%%%%%%%%%%%%%%%%%%%%%%%%%%%%%%%%%%%%%%%%%
\begin{figure}
%\vspace*{-3.cm}
\begin{center}
\begin{tabular}{@{\hspace{2.7cm}}c}
\epsfxsize17.0cm\epsffile{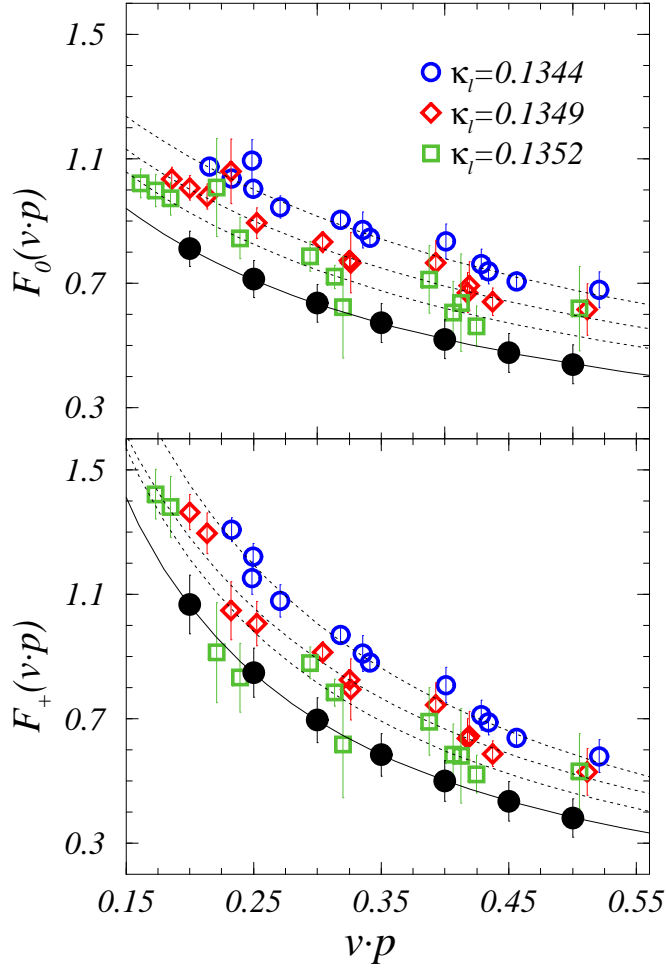}   \\
\end{tabular}
%%%%%%%%%%%%%%%%%%%%%%%%%%%%%%%%%%%%%%%%%%%%%%%%%%%%%%%%%%%%%%%%%%
\vspace*{-1.25cm}
\caption{\label{fig3}{\small Lattice data for the form factors 
$F_{0/+}(v\cdot p)$ with the heavy quark fixed to $\kappa_Q=0.1220$, and for 
three different values of light quark masses. Dotted (full) curves represent the fit to
the parameterization~(\ref{BK}). Filled symbols denote the form factors extrapolated
to the final pion at a given value of $v\cdot p$ which has been fixed according to eq.~(\ref{fixvp}).}}
%%%%%%%%%%%%%%%%%%%%%%%%%%%%%%%%%%%%%%%%%%%%%%%%%%%%%%%%%%%%%%%%%%
\end{center}
\end{figure}
%%%%%%%%%%%%%%%%%%%%%%%%%%%%%%%%%%%%%%%%%%%%%%%%%%%%%%%%%%%
We note that the lattice results for the form factors are 
in the region $0.2 \leq v\cdot p\leq 0.5$. In this window we could fix 7 equidistant points $v\cdot p = 0.20, 0.25,\dots 0.50$, where the
distance between the points is taken to be the same as in ref.~\cite{UKQCD}. 
We used the fit to the parameterization~(\ref{BK})~\footnote{Other interpolation forms are also 
applicable, such as the pole formula (with the pole mass left as a 
free parameter), which is what has been done in ref.~\cite{UKQCD}. We checked that the final results
on the form factors do not depend on the form used in this interpolation.} to interpolate to the
chosen values of $v\cdot p$. 
For each fixed $v\cdot p$, we then extrapolate to the final pion as described in the previous subsection. 
A typical situation is displayed in fig.~\ref{fig3}, where the heavy quark corresponds to
$\kappa_Q=0.1220$, {\it i.e.} $M_{H_d}=0.739(2)$ in lattice units. The result of the extrapolation is 
marked by the filled circles in fig.~\ref{fig3}. 

\begin{figure}
\vspace*{-7.7cm}
\begin{center}
\begin{tabular}{@{\hspace{-1.cm}}c}
\epsfxsize17.0cm\epsffile{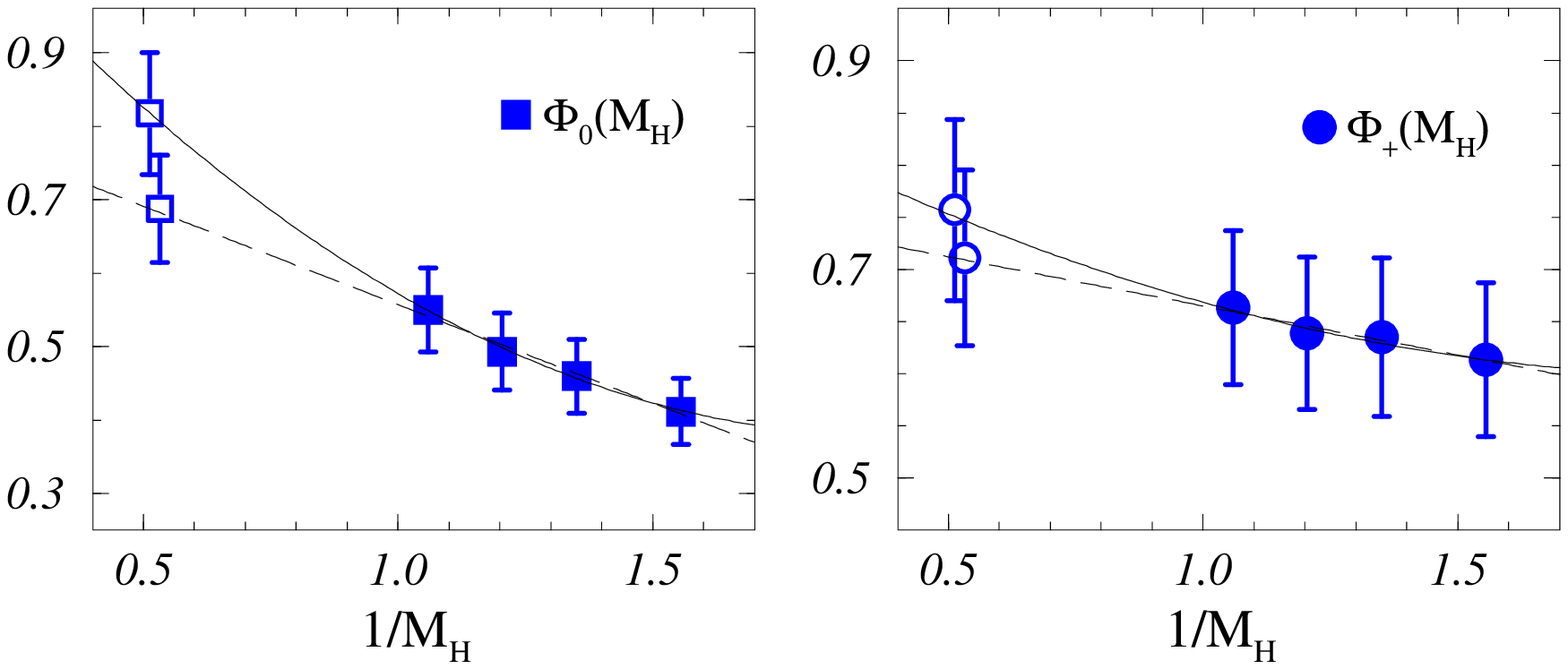}   \\
\end{tabular}
%%%%%%%%%%%%%%%%%%%%%%%%%%%%%%%%%%%%%%%%%%%%%%%%%%%%%%%%%%%%%%%%%%
\vspace*{-1.25cm}
\caption{\label{fig44}{\small Extrapolation of the form factors~(\ref{extra}) at $v\cdot p = 0.35$ in 
the inverse heavy meson mass to the $B_d$ meson. Dashed and full curves correspond to the linear and
quadratic extrapolation, respectively.}}
%%%%%%%%%%%%%%%%%%%%%%%%%%%%%%%%%%%%%%%%%%%%%%%%%%%%%%%%%%%%%%%%%%
\end{center}
\end{figure}

We now have to extrapolate in the heavy quark mass for which one uses the
heavy quark scaling laws. For each fixed value of $v\cdot p$, we extrapolate
\bea \label{extra}
\Phi(M_H)\ =\ a_0\ +\ {a_1\over \ M_H\ }\ +\  {a_2\over \ M_H\ } \;,
\eea
where
\bea \label{logg}
\Phi(M_H) 
=  \left( {\alpha_s(M_B)\over \alpha_s(M_H)}\right)^{-\widetilde \gamma_0/2\beta_0} \times 
\Bigl\{ \ F_0(v\cdot p) \sqrt{\ M_H\ }\  ,\; {F_+(v\cdot p) \over \sqrt{\ M_H\ }} \ \Bigr\} 
\eea
The term multiplying both form factors gives the logarithmic dependence 
on the heavy meson mass predicted by the HQET in which the vector
current has the anomalous dimension whose leading order coefficient is $\widetilde
\gamma_0=-4$~\cite{Grozin}~\footnote{It should be noted, however, that the effect of the 
logarithmic corrections in eq.~(\ref{logg}) on the extrapolation to $m_B$
is practically negligible.}.
As usual, the extrapolation in $1/M_H$~(\ref{extra}) is performed either linearly ($a_2=0$), 
or quadratically ($a_2$ free).
Obviously, the HQET scaling law~\cite{Isgur} is valid in the region close to the zero-recoil 
($v\cdot p\to 0$), and the extrapolation from the points which are rather far from zero-recoil is an
assumption. The difference between the results obtained after the linear and quadratic 
extrapolations will be included in the systematic uncertainty. 
In fig.~\ref{fig44}, we show this extrapolation for the case of $v\cdot p = 0.35$. 
%%%%%%%%%%%%%%%%%%%%%%%%%%%%%%%%%%%%%%%%%%%%%%%%%%%%%%%%%%%
%%%%%%%%%%%%%%%%%%%%%%%%%%%%%%%%%%%%%%%%%%%%%%%%%%%%%%%%%%%
%%%%%%%%%%%%%%%%%%%%%%%%%%%%%%%%%%%%%%%%%%%%%%%%%%%%%%%%%%%
%%%%%%%%%%%%%%%%%%%%%%%%%%%%%%%%%%%%%%%%%%%%%%%%%%%%%%%%%%%
As in the previous subsection we quote the result of the quadratic extrapolation
as our central value, and the difference with the linearly extrapolated value is included
in the estimate of the systematic error. 
The resulting form factors for the $B\to \pi$ semileptonic decay are given in tab.~\ref{tab:3}.
\begin{table}
\vspace*{-5mm}
\begin{center}
\begin{tabular}{|c|c|c|} 
  \hline
\hspace{-4.mm}{\phantom{\huge{l}}}\raisebox{-.2cm}{\phantom{\Huge{j}}}
{  $q^2\ [{\rm GeV}^2]$\hspace{3mm}}& { \hspace{-1mm}$F_0^{B\to \pi}(q^2)$\hspace{1mm}} &{\, \ $F_+^{B\to \pi}(q^2)$}\\ 
\hline   \hline 
{\phantom{\Large{l}}}\raisebox{.2cm}{\phantom{\Large{j}}}
{ 13.6} & {$\mathsf{0.46(7)^{+.05}_{-.08}}$} & {$\mathsf{0.70(9)^{+.10}_{-.03}}$} \\
{\phantom{\Large{l}}}\raisebox{.2cm}{\phantom{\Large{j}}}
{ 15.0} & {$\mathsf{0.49(7)^{+.06}_{-.08}}$} & {$\mathsf{0.79(10)^{+.10}_{-.04}}$} \\
{\phantom{\Large{l}}}\raisebox{.2cm}{\phantom{\Large{j}}}
{ 16.4} & {$\mathsf{0.54(6)^{+.05}_{-.09}}$} & {$\mathsf{0.90(10)^{+.10}_{-.04}}$} \\
{\phantom{\Large{l}}}\raisebox{.2cm}{\phantom{\Large{j}}}
{ 17.9} & {$\mathsf{0.59(6)^{+.04}_{-.10}}$} & {$\mathsf{1.05(11)^{+.10}_{-.06}}$} \\
{\phantom{\Large{l}}}\raisebox{.2cm}{\phantom{\Large{j}}}
{ 19.3} & {$\mathsf{0.64(6)^{+.04}_{-.10}}$} & {$\mathsf{1.25(13)^{+.09}_{-.08}}$} \\
{\phantom{\Large{l}}}\raisebox{.2cm}{\phantom{\Large{j}}}
{ 20.7} & {$\mathsf{0.71(6)^{+.03}_{-.10}}$} & {$\mathsf{1.53(17)^{+.08}_{-.11}}$} \\
{\phantom{\Large{l}}}\raisebox{.2cm}{\phantom{\Large{j}}}
{ 22.1} & {$\mathsf{0.80(6)^{+.01}_{-.12}}$} & {$\mathsf{1.96(23)^{+.06}_{-.18}}$} \\
  \hline 
\end{tabular}
%%%%%%%%%%%%%%%%%%%%%%%%%%%%%%%%%%%%%%%%%%%%%%%%%%%%%%%%%%%%%%%%%%
%\vspace*{-.75cm}
\caption{\label{tab:3}{\small The values of the semileptonic $B\to \pi$ form factors at several 
values of $q^2$. First errors are the statistical and the second are systematic whose estimate is
explained in the text.}}
\end{center}
\vspace*{-5mm}
\end{table}
The last step is the fit of the data from tab.~\ref{tab:3} to the form~(\ref{BK}), which is what we
discuss next.

\section{Physical results \label{Results}} 

\subsection{$B\to \pi\ell \nu_\ell$\label{ResultsB}}

\begin{figure}
\vspace*{-4.cm}
\begin{center}
\begin{tabular}{@{\hspace{.1cm}}c}
\epsfxsize16.8cm\epsffile{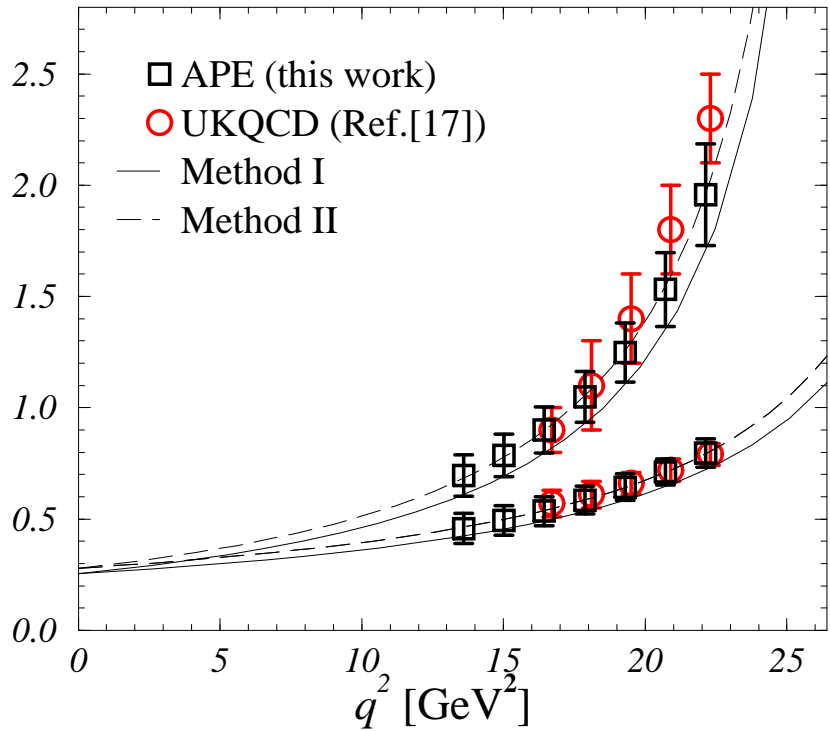}   \\
\end{tabular}
%%%%%%%%%%%%%%%%%%%%%%%%%%%%%%%%%%%%%%%%%%%%%%%%%%%%%%%%%%%%%%%%%%
\vspace*{-.95cm}
\caption{\label{fig8}{\small The form factors relevant for $B\to \pi \ell \bar \nu$ decay
 obtained by using {\sl Method II}, are directly compared to the result of the
 UKQCD group (only statistical errors are shown). We also draw the curves 
 describing the $q^2$ dependence with the parameters given in tab.~\ref{Bresults}. }}
%%%%%%%%%%%%%%%%%%%%%%%%%%%%%%%%%%%%%%%%%%%%%%%%%%%%%%%%%%%%%%%%%%
\end{center}
\end{figure}

The main physical results as obtained by employing the {\sl Method I} and  {\sl Method II}, 
are listed
in tab.~\ref{Bresults}.
\begin{table}
\vspace*{-.1cm}
\begin{center}
\hspace*{-3mm}\begin{tabular}{|c|ccc|} 
  \hline
\hspace*{-4mm}{\phantom{\huge{l}}}\raisebox{-.2cm}{\phantom{\Huge{j}}}
{\sl Quantity} & {\sl Method I}&  {\sl Method II}&  {\sl LCSR}~\cite{Khodjamirian}\\ 
\hline   
{\phantom{\Large{l}}}\raisebox{.2cm}{\phantom{\Large{j}}}

\hspace*{-4mm}$c_B$ &  $0.42(13)(4)$& $0.51(8)(1)$ & $0.41(12)$\\
{\phantom{\Large{l}}}\raisebox{.2cm}{\phantom{\Large{j}}}

\hspace*{-4mm}$\alpha_B$ &  $0.40(15)(9)$& $0.45(17)^{+.06}_{-.13}$ & $0.32^{+.21}_{-.07}$\\
{\phantom{\Large{l}}}\raisebox{.2cm}{\phantom{\Large{j}}}

\hspace*{-4mm}$\beta_B$ & $1.22(14)^{+.12}_{-.03}$& $1.20(13)^{+.15}_{-.00}$& --- \\ \hline 
{\phantom{\Large{l}}}\raisebox{.2cm}{\phantom{\Large{j}}}

\hspace*{-4mm}$F^{B\to \pi}(0)$ &  $0.26(5)(4)$ & $0.28(6)(5)$&
 $0.28(5)$\\
{\phantom{\Large{l}}}\raisebox{.2cm}{\phantom{\Large{j}}}

\hspace*{-4mm}$F^{B\to \pi}_0(m_B^2)$ & $1.3(6)^{+.0}_{-.4}$& $1.5(5)^{+.0}_{-.4}$ & --- \\
{\phantom{\Large{l}}}\raisebox{.2cm}{\phantom{\Large{j}}}
\hspace*{-4mm}$F^{B\to K}(0)/F^{B\to \pi}(0)$ & $1.21(9)^{+.00}_{-.09}$&
$1.19(11)^{+.03}_{-.11}$& $1.28^{+.18}_{-.10}$ \\
  \hline 
{\phantom{\Large{l}}}\raisebox{.2cm}{\phantom{\Large{j}}}
\hspace*{-4mm}$\vert V_{ub}\vert^{-2}\cdot \Gamma (\bar B^0\to \pi^+ \ell \bar \nu)\ [{\rm ps}^{-1}]$
&  $6.3\pm 2.4\pm 1.6$ & $8.5\pm 3.8\pm 2.2$ & $7.3\pm 2.5$
\\
{\phantom{\Large{l}}}\raisebox{.2cm}{\phantom{\Large{j}}}
\hspace*{-4mm}$\vert V_{ub}\vert /10^{-3}$ & 
$4.3\pm 0.9\pm 0.6$ & $4.0\pm 1.0\pm 0.5$  & 
$4.0\pm 0.7\pm 0.7$
\\
  \hline 
\end{tabular}
%%%%%%%%%%%%%%%%%%%%%%%%%%%%%%%%%%%%%%%%%%%%%%%%%%%%%%%%%%%%%%%%%%
\caption{\label{Bresults}{\small Main results of this paper, as obtained by using the two
methods explained in the text. For easier comparison, we also list the results obtained by
using the LCSR~{\rm \cite{Khodjamirian}}.}}
\end{center}
\end{table}
%%%%%%%%%%%%%%%%%%%%%%%%%%%%%%%%%%%%%%%%%%%%%%%%%%%%%%%%%%%%%%%%%%
%%%%%%%%%%%%%%%%%%%%%%%%%%%%%%%%%%%%%%%%%%%%%%%%%%%%%%%%%%%%%%%%%%
In the same table we also give the results obtained by using the LCSR which are fitted to the
parameterization~(\ref{BK}) in ref.~\cite{Khodjamirian}. First, we see that the final results 
of the two methods agree very well with each other. We also notice a very good agreement with the
results reported in ref.~\cite{UKQCD} where the {\sl Method II} has been used. This is 
shown in fig.~\ref{fig8}. Moreover, we see that there is a pleasant agreement with the LCSR results.
We verified that the Callan-Treiman relation~(\ref{SPT}) is well satisfied when the
quadratic extrapolation in the heavy meson mass is performed. We recall that on the same set of
configurations, we
have $f_B/f_\pi=1.29(10)$ and $1.46(16)$, as obtained from the linear, and quadratic extrapolation 
of the ratio $f_{H}/f_\pi$ in $1/M_H$, respectively. The latter value has to be compared with the value of
$F_0(m_B^2)$ given in tab.~\ref{Bresults}.

We were also able to examine the effect of the SU(3) breaking at $q^2=0$, by replacing the final pion
by a kaon. This ratio (see tab.~\ref{Bresults}) turns out to be large compared to one and, although with large
errors, quite compatible with the findings of the LCSR.
%%%%%%%%%%%%%%%%%%%%%%%%%%%%%%%%%%%%%%%%%%%%%%%%%%%%%%%%%%%
%%%%%%%%%%%%%%%%%%%%%%%%%%%%%%%%%%%%%%%%%%%%%%%%%%%%%%%%%%%
%%%%%%%%%%%%%%%%%%%%%%%%%%%%%%%%%%%%%%%%%%%%%%%%%%%%%%%%%%%
%%%%%%%%%%%%%%%%%%%%%%%%%%%%%%%%%%%%%%%%%%%%%%%%%%%%%%%%%%%

We now discuss the sources of systematic errors which are combined in the quadrature.
\begin{itemize}

\item[--] {\sf Extraction of the matrix element:} The central results are obtained
by using the {\sl ratio method} (see sec.~\ref{Lattice}). The whole analysis has also 
been performed with the matrix element extracted by using the {\sl analytic method}. 
The difference in the results is always smaller than the statistical error, and we 
include it in the systematic uncertainty.

\item[--] {\sf Discretization errors:} The main source of discretization errors is
expected to be of ${\cal O}( (am_H)^2)$, where $m_H$ is the heavy meson mass.
Our central results are obtained by correcting the 
${\cal O}(a)$ improved renormalization constant by the KLM factor as given in eq.~(\ref{klm}).
We repeated the analysis without the KLM factor and found only a small discrepancy 
with the central results of the parameter $c_B$ (of about $5~\%$) which is included in the
systematics. The other parameters remain unchanged. We emphasize that this comparison can only 
give us idea about the size of discretization errors. 
To check this point we computed 
$Z_V$ for the case 
of the heavy-heavy vector current $\bar Q\gamma_0 Q$ for each of
our heavy quarks. By considering the ratio
\bea
\label{zv}
{\cal R} (t) &=& 
{
{\displaystyle \int} d \vec x d \vec y\ \langle   H(y)\ \bar Q\gamma_0 Q({\vec x}, t)\ H^\dagger(0)  \rangle\, \over
2\ {\displaystyle \int} d \vec x \ \langle   H({\vec x}, t) 
H^{\dagger}(0) \rangle\,},
\eea
which  
is shown in fig.~\ref{fig66}, one directly accedes $1/Z_V$. We follow ref.~\cite{UKQCD}, and denote such an obtained $Z_V$ as 
$Z_V^{(eff)}$ which we now compare to the values that one gets by using the expression 
$Z_V=Z_V^{(0)}(1 + b_V am_Q)$, where $Z_V^{(0)}$ and $b_V$ are those given in tab.~\ref{tab:1}: 
\bea
\kappa_Q=0.1250;&& Z_V^{(eff)} = 1.155(31)\;\; [Z_V = 1.144(6)]\,,\cr
&&\cr
\kappa_Q=0.1220;&& Z_V^{(eff)} = 1.254(35)\;\; [Z_V = 1.252(6)]\,,\cr
&&\cr
\kappa_Q=0.1190;&& Z_V^{(eff)} = 1.363(40)\;\; [Z_V = 1.367(6)]\,,\cr
&&\cr
\kappa_Q=0.1150;&& Z_V^{(eff)} = 1.516(47)\; [Z_V = 1.528(6)].
\eea
Clearly, results of the two procedures are fully consistent
with each other. In other words, there is no room for a term proportional to $(a m_Q)^2$.~\footnote{From the fit 
$Z_V^{(eff)}=Z_V^{(0)}(1 + b_V am_Q + b_V^{(eff)} (am_Q)^2)$, where the coefficients $Z_V^{(0)}$ and $b_V$ are given in tab.~\ref{tab:1}, we obtain $b_V^{(eff)}=0.05(27)$, thus consistent with zero.} 
%%%%%%%%%%%%%%%%%%%%%%%%%%%%%%%%%%%%%%%%%%%%%%%%%%%%%%%%%%%
\begin{figure}[h!!]
%\vspace*{-2.cm}
\begin{center}
\begin{tabular}{@{\hspace{.1cm}}c}
\epsfxsize10.0cm\epsffile{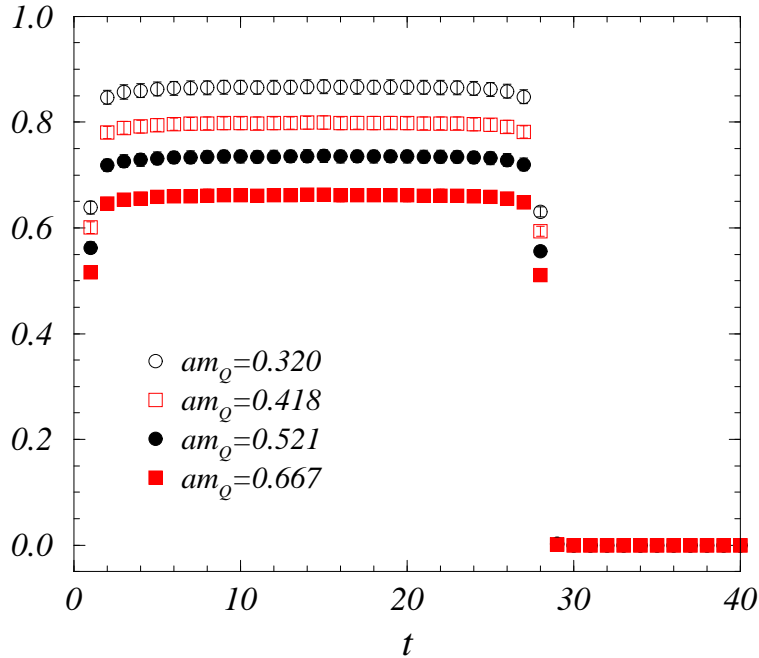}   \\
\end{tabular}
%%%%%%%%%%%%%%%%%%%%%%%%%%%%%%%%%%%%%%%%%%%%%%%%%%%%%%%%%%%%%%%%%%
\vspace*{-.5cm}
\caption{\label{fig66}{\small $1/Z_V^{(eff)}$ as obtained from 
the ratio ${\cal R} (t)$ (see eq.~(\ref{zv})) for the time 
slices between the two source operators  
(placed at $0$ and $27$ respectively). Ratio is plotted for 
each heavy quark mass $2 m_Q=1/\kappa_Q - 1/\kappa_{cr}$. Fit is performed for $t\in [5,25]$.}}
%%%%%%%%%%%%%%%%%%%%%%%%%%%%%%%%%%%%%%%%%%%%%%%%%%%%%%%%%%%%%%%%%
\end{center}
\end{figure}

Besides, we also used the value of $a^{-1}(f_K)=2.8(1)$~GeV and repeated the whole analysis which 
leads to the results only slightly different (less than $5\%$) from our central values.
This difference is, however, added in the systematic uncertainty.

\item[--] {\sf Heavy meson extrapolation:} As already mentioned, we quote as central values the results
obtained by using the quadratic extrapolation in $1/M_H$. The difference between this and the results 
of the linear extrapolation is added to the systematics. Beside quenching, this is the main source of
systematic uncertainty present in our analysis (it is of order $15\%$).

\end{itemize}

\begin{figure}
\vspace*{-5.cm}
\begin{center}
\begin{tabular}{@{\hspace{.1cm}}c}
\epsfxsize16.0cm\epsffile{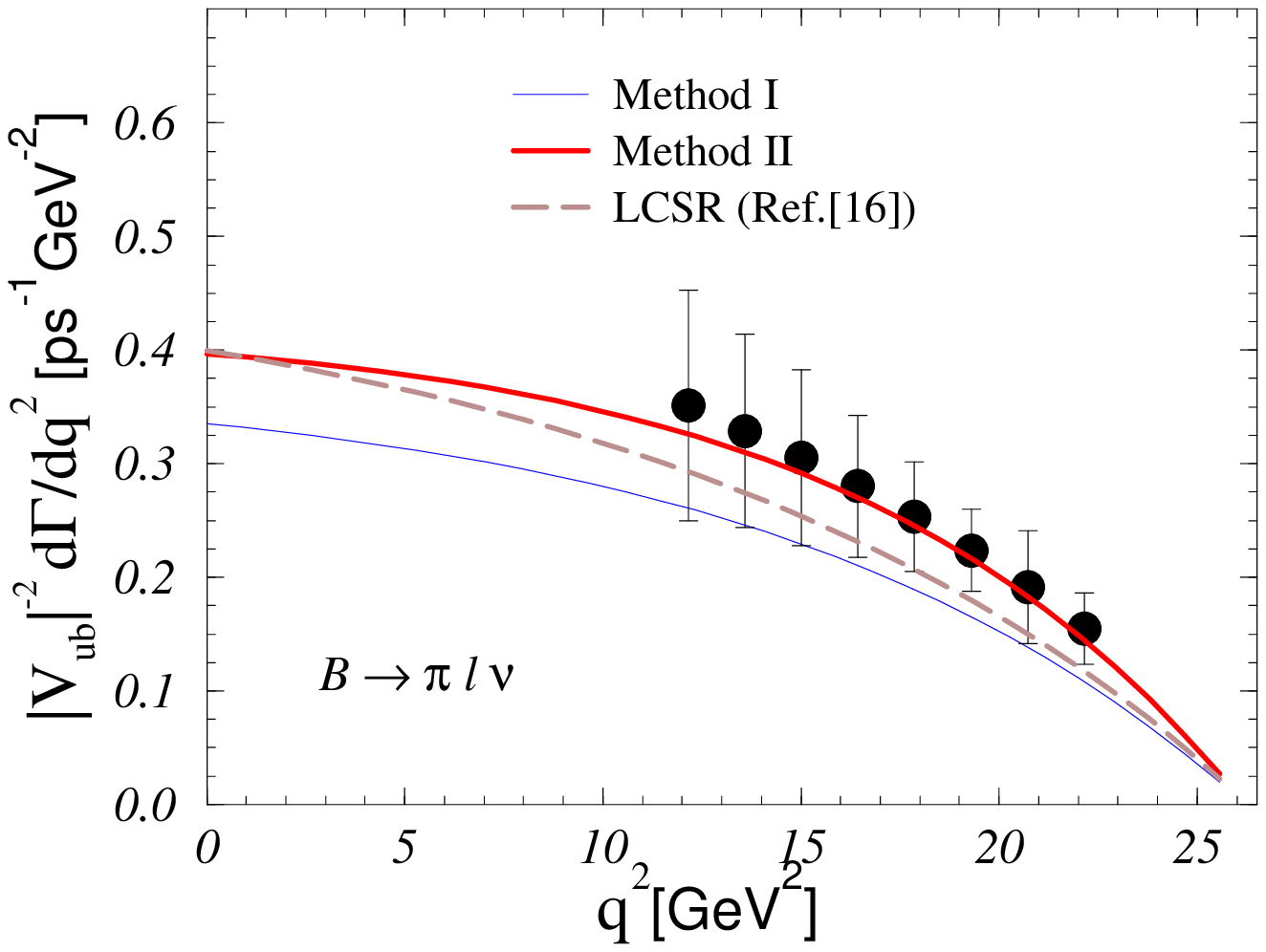}   \\
\end{tabular}
%%%%%%%%%%%%%%%%%%%%%%%%%%%%%%%%%%%%%%%%%%%%%%%%%%%%%%%%%%%%%%%%%%
%\vspace*{-1.25cm}
\caption{\label{fig9}{\small Differential decay rate for $B\to \pi \ell \bar \nu$ decay. We show the
curves as obtained by both methods. We also plot the result of ref.~\cite{Khodjamirian}, 
obtained using the LCSR.}}
%%%%%%%%%%%%%%%%%%%%%%%%%%%%%%%%%%%%%%%%%%%%%%%%%%%%%%%%%%%%%%%%%%
\end{center}
\end{figure}

The small differences of BK-parameters given in tab.~\ref{Bresults} give a more 
pronounced effect in the $q^2$ spectrum of the differential decay rate~(\ref{rate}). 
This is shown in fig.~\ref{fig9}.
%%%%%%%%%%%%%%%%%%%%%%%%%%%%%%%%%%%%%%%%%%%%%%%%%%%%%%%%%%%
%%%%%%%%%%%%%%%%%%%%%%%%%%%%%%%%%%%%%%%%%%%%%%%%%%%%%%%%%%%
After integrating eq.~(\ref{rate}) we get the decay width (divided by 
$\vert V_{ub}\vert^2$), which is also given in tab.~\ref{Bresults}. As mentioned in the
introduction, after comparing this result to the measured branching ratio $B(\bar B^0\to \pi^+
\ell \bar \nu )$~\cite{CLEO}, we are able to make an estimate of $\vert V_{ub}\vert$~\footnote{ 
The last
number, however, should be taken cautiously because the experimental result has been 
obtained by relying on various 
different theoretic estimates.}. We also checked that if the pole/dipole form is used to describe 
the $q^2$-dependence of the form factors that we obtained by using {\sl Method II}, 
the result is fully consistent with the one
we quote in tab.~\ref{Bresults}, namely we obtain 
$\vert V_{ub}\vert^{-2}\cdot \Gamma (\bar B^0\to \pi^+ \ell \bar \nu) = 
7.8\pm 3.2\pm 2.1\ {\rm ps}^{-1}$.

To make a unique estimate, we combined in  quadrature the statistical and systematic 
errors for $\vert V_{ub}\vert^{-2}\ \Gamma (\bar B^0\to \pi^+ \ell \bar \nu)$, and made a weighted
average of the results of two methods to get, 
$\vert V_{ub}\vert^{-2}\ \Gamma (\bar B^0\to \pi^+ \ell \bar \nu)= (7.0\pm 2.9)$~ps. Now, as before,
if we
compare this result to the measured branching ratio, we obtain
$\vert V_{ub}\vert = (4.1 \pm 1.1)\cdot 10^{-3}$, which is consistent with the world averaged 
value from inclusive decays, presented at this year's ICHEP conference~\cite{ICHEP}: 
$\vert V_{ub}\vert = (4.1 \pm 0.6\pm 0.2)\cdot 10^{-3}$.

\section{$D\to \pi\ell \nu_\ell$ and $D\to K\ell \nu_\ell$ \label{ResultsD}}

\begin{figure}
\vspace*{-3.cm}
\begin{center}
\begin{tabular}{@{\hspace{-.7cm}}c}
\epsfxsize15.6cm\epsffile{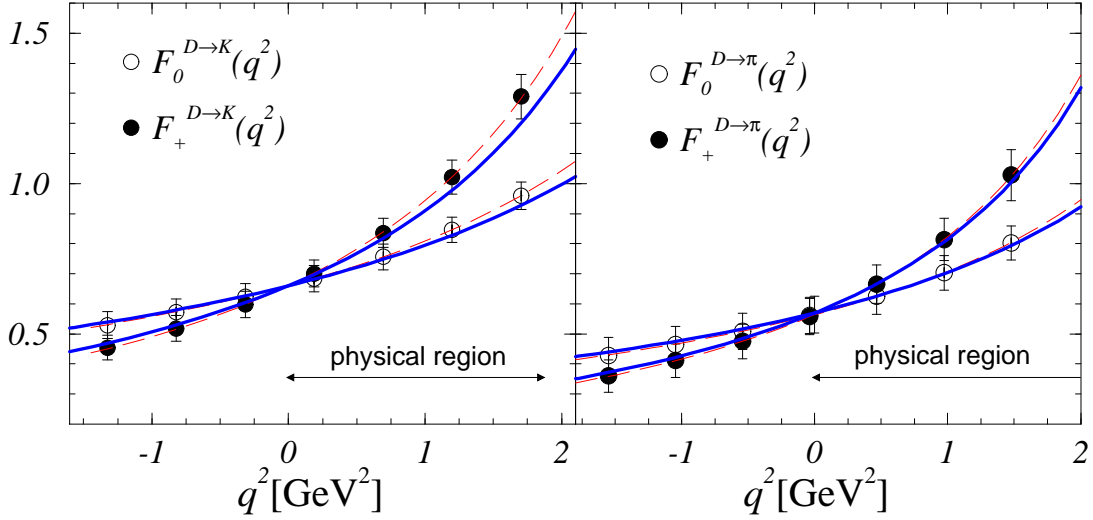}   \\
\end{tabular}
%%%%%%%%%%%%%%%%%%%%%%%%%%%%%%%%%%%%%%%%%%%%%%%%%%%%%%%%%%%%%%%%%%
\vspace*{-1.05cm}
\caption{\label{fig4}{\small Form factors for $D\to K$ and $D\to \pi$ semileptonic decays. 
The full line correspond to the results of {\sl Method I}, and the dashed one to the results
of {\sl Method II}. The physical region $0\leq q^2\leq (m_D-m_{K/\pi})^2$ 
is indicated in both cases.}}
%%%%%%%%%%%%%%%%%%%%%%%%%%%%%%%%%%%%%%%%%%%%%%%%%%%%%%%%%%%%%%%%%%
\end{center}
\end{figure}

In this section we briefly discuss the results relevant for the semileptonic decays
of $D$ meson. As in the previous section, we apply both {\sl Method I} and {\sl Method II}. In this case,
however, the most important source of systematic uncertainty is negligible, namely 
in order to reach the physical value of the heavy meson mass, only a smooth interpolation of 
the data from tab.~\ref{tab:4}, by using eq.~(\ref{hqs}), is required.
Notice that for these decays, a bulk of final lattice points is out of the
physical region (this is illustrated in fig.~\ref{fig4}).
The values of the form factors whose $q^2$'s are within the physical region for the semileptonic decays  
are listed in tab.~\ref{tab:8}.

Our final results are given in tab.~\ref{Dresults}, where we again compare to the numbers obtained 
from the LCSR study of ref.~\cite{Khodjamirian}. 

%%%%%%%%%%%%%%%%%%%%%%%%%%%%%%%%%%%%%%%%%%%%%%%%%%%%%%%%%%%
%%%%%%%%%%%%%%%%%%%%%%%%%%%%%%%%%%%%%%%%%%%%%%%%%%%%%%%%%%%
%%%%%%%%%%%%%%%%%%%%%%%%%%%%%%%%%%%%%%%%%%%%%%%%%%%%%%%%%%%
\begin{table}
%%%%%%%%%%%%%%%%%%%%%%%%%%%%%%%%%%%%%%%%%%%%%%%%%%%%%%%%%%%
%\vspace*{-.75cm}
\begin{center}
\begin{tabular}{|c||c|c|c|c|} 
  \hline
{\phantom{\huge{l}}}\raisebox{-.2cm}{\phantom{\Huge{j}}}
{  $q^2\ [{\rm GeV}^2]$}& {-0.04} &  {0.47}  &  {0.97} &{ 1.48}  \\ \hline
{\phantom{\Large{l}}}\raisebox{.2cm}{\phantom{\Large{j}}}

{\phantom{\Large{l}}}\raisebox{+.2cm}{\phantom{\Large{j}}}
{ \hspace{-1mm}$F_0^{D\to \pi}(q^2)$\hspace{1mm}} & 
{$\mathsf{0.56(6)^{+.02}_{-.00}}$} & 
{$\mathsf{0.62(6)^{+.02}_{-.00}}$} & 
{$\mathsf{0.70(6)^{+.01}_{-.00}}$} & 
{$\mathsf{0.80(6)^{+.01}_{-.00}}$}  \\    
{\phantom{\Large{l}}}\raisebox{-.2cm}{\phantom{\Large{j}}}
{\, \ $F_+^{D\to \pi}(q^2)$}  & 
{$\mathsf{0.56(6)^{+.02}_{-.00}}$} & 
{$\mathsf{0.67(6)^{+.01}_{-.00}}$} & 
{$\mathsf{0.81(7)^{+.02}_{-.00}}$} & 
{$\mathsf{1.03(9)^{+.01}_{-.00}}$}  \\    
  \hline \hline
{\phantom{\huge{l}}}\raisebox{-.2cm}{\phantom{\Huge{j}}}
{  $q^2\ [{\rm GeV}^2]$}& {0.19} &  {0.69}  &  {1.20} &{ 1.70}   \\ \hline
{\phantom{\Large{l}}}\raisebox{.2cm}{\phantom{\Large{j}}}

{\phantom{\Large{l}}}\raisebox{+.2cm}{\phantom{\Large{j}}}
{ \hspace{-1mm}$F_0^{D\to K}(q^2)$\hspace{1mm}} & 
{$\mathsf{0.68(4)(0)}$} & 
{$\mathsf{0.76(4)(0)}$} & 
{$\mathsf{0.85(4)(0)}$} & 
{$\mathsf{0.96(4)(0)}$}  \\    
{\phantom{\Large{l}}}\raisebox{-.2cm}{\phantom{\Large{j}}}
{\, \ $F_+^{D\to K}(q^2)$}  & 
{$\mathsf{0.70(5)(0)}$} & 
{$\mathsf{0.84(5)(0)}$} & 
{$\mathsf{1.02(6)(0)}$} & 
{$\mathsf{1.29(7)(0)}$}  \\    
  \hline  
\end{tabular}
%%%%%%%%%%%%%%%%%%%%%%%%%%%%%%%%%%%%%%%%%%%%%%%%%%%%%%%%%%%%%%%%%%
%\vspace*{-1.25cm}
\caption{\label{tab:8}{\small The semileptonic $D\to \pi$ and $D\to K$ form factors 
at several $q^2$'s obtained by using {\sl Method II}. Only the values at $q^2$ belonging 
to the physical region for the semileptonic decays are
listed, with the exception of $q^2=-0.04\ \gev^2$ which is very close to $q^2=0$.
The systematic errors are estimated in the way explained in the text.}}
\end{center}
\vspace*{-.25cm}
\end{table}

\begin{figure}
%\vspace*{-3.cm}
\begin{center}
\begin{tabular}{@{\hspace{.1cm}}c}
\epsfxsize17.0cm\epsffile{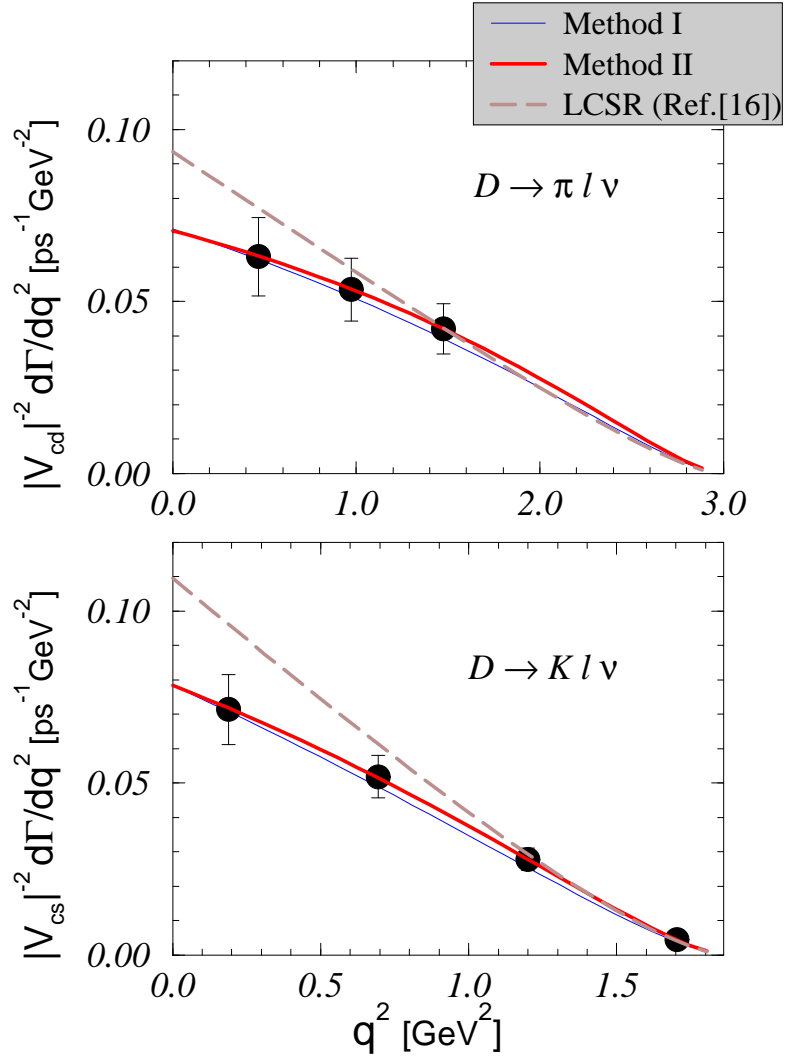}   \\
\end{tabular}
%%%%%%%%%%%%%%%%%%%%%%%%%%%%%%%%%%%%%%%%%%%%%%%%%%%%%%%%%%%%%%%%%%
%\vspace*{-1.25cm}
\caption{\label{fig10}{\small The same as in fig.~(\ref{fig9}) but for
$D$ meson semileptonic decays.}}
%%%%%%%%%%%%%%%%%%%%%%%%%%%%%%%%%%%%%%%%%%%%%%%%%%%%%%%%%%%%%%%%%%
\end{center}
\end{figure}
\begin{table}
\vspace*{-.1cm}
\begin{center}
\hspace*{-3mm}\begin{tabular}{|c|ccc|} 
  \hline
\hspace*{-4mm}{\phantom{\huge{l}}}\raisebox{-.2cm}{\phantom{\Huge{j}}}
{\sl Quantity} & {\sl Method I}&  {\sl Method II}&  {\sl LCSR}~\cite{Khodjamirian}\\ 
\hline   
{\phantom{\Large{l}}}\raisebox{.2cm}{\phantom{\Large{j}}}

\hspace*{-4mm}$c_{DK}$ &  $0.90(12)$& $1.17(23)$ & ---\\
{\phantom{\Large{l}}}\raisebox{.2cm}{\phantom{\Large{j}}}

\hspace*{-4mm}$\alpha_{DK}$ &  $0.27(11)^{+.00}_{-.01}$& $0.43(12)^{+.00}_{-.02}$ & $-0.07^{+.15}_{-.07}$\\
{\phantom{\Large{l}}}\raisebox{.2cm}{\phantom{\Large{j}}}

\hspace*{-4mm}$\beta_{DK}$ & $1.34(14)^{+.05}_{-.00}$& $1.22(13)^{+.05}_{-.00}$& --- \\ \hline 
{\phantom{\Large{l}}}\raisebox{.2cm}{\phantom{\Large{j}}}

\hspace*{-4mm}$c_{D\pi}$ &  $0.76(13)^{+.01}_{-.06}$& $0.88(18)^{+.00}_{-.06}$ & ---\\
{\phantom{\Large{l}}}\raisebox{.2cm}{\phantom{\Large{j}}}

\hspace*{-4mm}$\alpha_{D\pi}$ &  $0.27(14)^{+.00}_{-.06}$& $0.36(16)^{+.00}_{-.07}$ & $0.01^{+.11}_{-.07}$\\
{\phantom{\Large{l}}}\raisebox{.2cm}{\phantom{\Large{j}}}

\hspace*{-4mm}$\beta_{D\pi}$ & $1.31(17)^{+.13}_{-.00}$& $1.23(17)^{+.15}_{-.00}$& --- \\ \hline 
{\phantom{\Large{l}}}\raisebox{.2cm}{\phantom{\Large{j}}}

\hspace*{-4mm}$F^{D\to K}(0)$ &  $0.66(4)^{+.01}_{-.00}$ & $0.66(4)^{+.01}_{-.00}$& $0.78(11)$\\
{\phantom{\Large{l}}}\raisebox{.2cm}{\phantom{\Large{j}}}

\hspace*{-4mm}$F^{D\to \pi}(0)$ &  $0.57(6)^{+.01}_{-.00}$ & $0.57(6)^{+.02}_{-.00}$& $0.65(11)$\\
{\phantom{\Large{l}}}\raisebox{.2cm}{\phantom{\Large{j}}}

\hspace*{-4mm}$F^{D\to \pi}_0(m_D^2)$ & $1.6(3)^{+.0}_{-.2}$& $1.9(5)^{+.0}_{-.3}$ & --- \\
{\phantom{\Large{l}}}\raisebox{.2cm}{\phantom{\Large{j}}}
\hspace*{-4mm}$F^{D\to K}(0)/F^{D\to \pi}(0)$ & $1.16(5)^{+.00}_{-.03}$& $1.22(6)$& $\simeq 1.2$ \\
  \hline 
\end{tabular}
%%%%%%%%%%%%%%%%%%%%%%%%%%%%%%%%%%%%%%%%%%%%%%%%%%%%%%%%%%%%%%%%%%
\caption{\label{Dresults}{\small The same as in tab.~\ref{Bresults}.
}}
\end{center}
\end{table}
%%%%%%%%%%%%%%%%%%%%%%%%%%%%%%%%%%%%%%%%%%%%%%%%%%%%%%%%%%%%%%%%%%
As in the previous section, we plot the $q^2$ dependence of the partial decay width for both
decays ($D\to K$ and $D\to \pi$). The agreement with the results of the LCSR~\footnote{
The quoted LCSR results for $D\to K \ell \nu$~\cite{Khodjamirian} are  
obtained by using $m_s(\mu=m_c)=150\ \mev$.}  
is less good than it was for $B\to \pi \ell \nu$. Our central 
numbers for the 
form factors at $q^2=0$ are smaller. This region (of small $q^2$'s) dominates  
when integrating over the phase space. On the other hand the discrepancy in the $q^2$-dependence of the form factors 
is less important since the kinematically accessible region from these decays is small. 
Our results for the parameter $\alpha$, larger than those obtained by LCSR, suggest 
that the contribution of the
higher excited states to the form factor $F_+(q^2)$ is not absent. LCSR, instead, predict that the $q^2$-dependence
of $F_+^{D\to \pi (K)}(q^2)$ is very close to a pure pole behavior 
(with the pole mass being $m_{D^*_{(s)}}$). 
We also mention that $c_D$ gives access to the value of the strong coupling of 
$D$ and $\pi$ to the first pole $D^*$ (known as $g_{D^*D\pi}$), namely $g_{D^*D\pi}= 2 c_D
m_{D^*}/f_{D^*}$. With $f_{D^*}=258(14)(6)$~MeV~\cite{Bangalore}, this amounts to
$g_{D^*D\pi} = 12\pm 2$ and $g_{D^*D\pi} = 14\pm 5$, by using the {\sl Method I} and 
{\sl Method II}, respectively. The most recent LCSR prediction is $g_{D^*D\pi}=10.5\pm
3$~\cite{Weinzierl}~\footnote{The coupling $g_{D^*D\pi}$ is one of the main parameters 
in the approach based on the phenomenological lagrangians~\cite{Casalbuoni}.}.
A detailed discussion on this coupling with a complete list of references can be found in refs.~\cite{Singer}.

\begin{table}
\vspace*{-.1cm}
\begin{center}
\hspace*{-3mm}\begin{tabular}{|c|c|c|} 
  \hline
\hspace*{-4mm}{\phantom{\huge{l}}}\raisebox{-.2cm}{\phantom{\Huge{j}}}
&$ \Gamma ( D^0\to K^- \ell \nu)\ [{\rm ps}^{-1}]$&  
$ \Gamma ( D^0\to \pi^- \ell \nu)\ [{\rm ps}^{-1}]$\\ \hline 
{\phantom{\Large{l}}}\raisebox{.2cm}{\phantom{\Large{j}}}
{\rm Experiment\;}{\rm \cite{PDG}}& $(8.5\pm 0.5)\cdot 10^{-2} $& 
$(8.6\pm 1.8)\cdot 10^{-3} $\\ \hline
{\phantom{\Large{l}}}\raisebox{.2cm}{\phantom{\Large{j}}}
{\rm This\ work\;} ({\sl Method I})& $(6.9\pm 1.1)\cdot 10^{-2}$ &  $(5.6\pm 1.4)\cdot 10^{-3}$\\ 
{\phantom{\Large{l}}}\raisebox{.2cm}{\phantom{\Large{j}}}
{\rm This\ work\;} ({\sl Method II})& $(7.3\pm 1.1)\cdot 10^{-2}$ & $(5.9\pm 1.5)\cdot 10^{-3}$ \\ \hline
{\phantom{\Large{l}}}\raisebox{.2cm}{\phantom{\Large{j}}}
{\rm APE 95}~\cite{crisafulli}& $(9.1\pm 2.0)\cdot 10^{-2}$ & $(8\pm 2)\cdot 10^{-3}$\\ 
{\phantom{\Large{l}}}\raisebox{.2cm}{\phantom{\Large{j}}}
{\rm UKQCD 95}~\cite{bowler10}& $(7.0\pm 1.7)\cdot 10^{-2}$ & $(5.2\pm 1.8)\cdot 10^{-3}$\\ 
{\phantom{\Large{l}}}\raisebox{.2cm}{\phantom{\Large{j}}}
{\rm LANL 95}~\cite{gupta96}& $(9.4\pm 1.5)\cdot 10^{-2}$ & $(7.4\pm 1.1)\cdot 10^{-3}$\\ \hline
{\phantom{\Large{l}}}\raisebox{.2cm}{\phantom{\Large{j}}}
{\rm LCSR}~\cite{Khodjamirian}& $(9.2\pm 3.5)\cdot 10^{-2}$ & $(6.3\pm 2.4)\cdot 10^{-3}$\\ 
  \hline 
\end{tabular}
%%%%%%%%%%%%%%%%%%%%%%%%%%%%%%%%%%%%%%%%%%%%%%%%%%%%%%%%%%%%%%%%%%
\caption{\label{Drates}{\small Comparison of the full decay widths obtained in this work,
to the experimental value, to the several recent lattice results and to the prediction 
of ref.~\cite{Khodjamirian}. In the 
computation we used $\vert V_{cs}\vert = 0.9745(8)$
and $\vert V_{cd}\vert = 0.220(3)$~\cite{PDG}.
}}
\end{center}
\end{table}
%%%%%%%%%%%%%%%%%%%%%%%%%%%%%%%%%%%%%%%%%%%%%%%%%%%%%%%%%%%%%%%%%%

Finally, in
tab.~\ref{Drates} we also give our results for the total decay rates, which are in a fair
agreement with the experimental values~\cite{PDG}. As in the previous subsection, we can also 
compare our result for $\vert V_{cs}\vert^{-2} \Gamma ( D^0\to \pi^- \ell \nu)$ to the experimental
measurement $\Gamma^{\rm (exp.)} ( D^0\to K^- \ell \bar \nu) = (8.5\pm 0.5)\cdot 10^{-2} {\rm ps}^{-1}$.
We obtain,
\bea
\vert V_{cs}\vert = 1.07\pm 0.09\;,
\eea
which is to be compared to $\vert V_{cs}\vert = 1.04\pm 0.16$~\cite{PDG}, extracted from the
$D\to K$ semileptonic decay where the vector meson dominance for the form 
factor $F_+(q^2)$ has been assumed.
Both central values are larger than $1$, although consistent with $\vert V_{cs}\vert =
0.9745(8)$~\cite{PDG}, the value obtained by using the unitarity of the CKM matrix. 
The more precise experimental result for $B(D^0\to K^- \ell \nu)$ expected from 
CLEO and E791~\cite{ICHEP} will hopefully help in clarifying the situation.

\section{Conclusion \label{Conclusion}}

In this paper we have computed the form factors relevant for the semileptonic decays of the heavy-to-light
pseudoscalar mesons on the lattice. We used the fully ${\cal O}(a)$ improved Wilson action and 
 vector current. Since the actual computation is performed on a lattice with an inverse lattice 
spacing smaller than
$3$~GeV, the physical results for the most interesting decay in this class, namely $B\to \pi \ell
\nu$, had to be reached through an extrapolation in the inverse heavy meson mass, for which 
HQET and LEET provide us helpful scaling laws. To describe 
the interplay of the mass dependence and the
$q^2$-dependence of the form factors, we used two different methods which are described 
in the text. We summarize our findings as follows:
\begin{itemize}
\item[--] There is a good agreement of the results obtained by using two different 
methods to reach the physically interesting results of our study;
\item[--] We note a very good agreement with the results of ref.~\cite{UKQCD} in which the same 
(improved) lattice action has been used;
\item[--] Our results agree with the ones of ref.~\cite{Khodjamirian} where the LCSR were employed.
This is especially the case for $B\to \pi \ell \nu$ decay.  We also note that, although 
with somewhat large errors, we see SU(3) breaking effects 
for the form factors at $q^2=0$, consistent with the findings 
of the LCSR studies;
\item[--] We also verified that our form factor data are consistent with 
the soft pion theorem (Callan-Treiman relation~(\ref{SPT})).
\end{itemize}
The main results of our study are given in the introduction 
and in tables \ref{Bresults}--\ref{Drates}. A remaining 
uncertainty for the $B\to \pi$ decay arises from our inability to
work with the $B$ meson directly on the lattice. This uncertainty can be
further reduced by repeating the calculation on a larger and finer
lattice. Besides, our lattice simulation has been performed in the quenched approximation. 
It is obviously desirable to attempt the unquenched study of the semileptonic decays.

\newpage
%\vspace*{30mm}

\noindent
\underline{\large{Acknowledgements}}

\vspace*{10mm}
It is a pleasure to thank G.~Martinelli for many interesting discussions, C.~Bernard, A.~Khodjamirian, 
L.~Lellouch and D.~Melikhov for their valuable comments on the manuscript. 
A.A thanks the CERN Theory Division, where this work has 
been partially done.
The financial support of M.U.R.S.T. and INFN are kindly acknowledged. 
Rome and Orsay teams are members of the
european network ``Hadron Phenomenology from Lattice QCD'', HPRN-CT-2000-00145.

%\newpage
\vspace*{30mm}
%%%%%%%%%%%%%  References

\end{document}